\documentclass[prb,amsmath,amssymb,superscriptaddress,twocolumn]{revtex4}
\usepackage{braket}
\usepackage{graphicx}
\usepackage{bbm,color,ulem}
\DeclareMathAlphabet{\mathpzc}{OT1}{pzc}{m}{it}
\begin{document}
\title{Environmental noise effects on entanglement fidelity of exchange-coupled semiconductor spin qubits}
\author{Robert E.\ Throckmorton}
\affiliation{Condensed Matter Theory Center and Joint Quantum Institute, Department of Physics, University of Maryland, College Park, Maryland 20742-4111 USA}
\author{Edwin Barnes}
\affiliation{Condensed Matter Theory Center and Joint Quantum Institute, Department of Physics, University of Maryland, College Park, Maryland 20742-4111 USA}
\affiliation{Department of Physics, Virginia Tech, Blacksburg, Virginia 24061, USA}
\author{S.\ Das Sarma}
\affiliation{Condensed Matter Theory Center and Joint Quantum Institute, Department of Physics, University of Maryland, College Park, Maryland 20742-4111 USA}
\date{\today}
\begin{abstract}
We investigate the effect of magnetic field and charge noise on the generation of
entanglement between two Heisenberg exchange-coupled electron spins in a double quantum dot. We focus
on exchange-driven evolution that would ideally take an initial unentangled tensor product state to a
maximally-entangled state in the absence of noise. The presence of noise obviously adversely affects
the attainment of maximal entanglement, which we study quantitatively and exactly. To quantify the
effects of noise, we calculate two-qubit coherence times and entanglement fidelity, both of which can
be extracted from simulations or measurements of the return probability as a function of interaction
time, i.e., the time period during which the exchange coupling remains effective between the two spins.
We perform these calculations for a broad range of noise strengths that includes the regime of recent
experiments. We find that the two types of noise reduce the amount of entanglement in qualitatively
distinct ways and that, although charge noise generally leads to faster decoherence, the relative importance
of the two types of noise in entanglement creation depends sensitively on the strength of the exchange
coupling.  Our results can be used to determine the level of noise suppression needed to reach quantum
error correction thresholds. We provide quantitative guidance for the requisite noise constraints necessary
to eventually reach the $>99\%$ fidelity consistent with the quantum error correction threshold.
\end{abstract}
\maketitle

\section{Introduction}
Electron spins in semiconductor quantum dots are an attractive platform for quantum computation for
two reasons.  One is the simple fact that this platform is compatible with the existing semiconductor
electronics industry, making scalability much more feasible\cite{ZwanenburgRMP2013}.  The other is the
possibility of faster operations compared to other platforms, such as trapped ions or atoms and qubits
based on superconducting circuits.  These advantages have provided the impetus behind the substantial
theoretical and experimental progress on GaAs- and Si-based systems that has been made over the past
decade. High-fidelity single-qubit gates and control over multiqubit arrays have been demonstrated
in several different types of spin qubits, including single-spin Loss-DiVincenzo qubits\cite{LossPRA1998,PlaNature2012,PlaNature2013,VeldhorstNatNano2014,BraakmanNatNano2013,OtsukaSciRep2016,ItoArxiv2016},
double-dot singlet-triplet qubits\cite{LevyPRL2002,PettaScience2005,FolettiNatPhys2009,MauneNature2012,ShulmanNatCommun2014,DialPRL2013,MartinsPRL2016},
triple-dot exchange-only qubits\cite{DiVincenzoNature2000,MedfordNatNano2013,MedfordPRL2013,EngSciAdv2015,ShimPRB2016}, and
hybrid qubits\cite{ShiPRL2012,KimNature2014,KimnpjQI2015} consisting of three electrons confined in a double quantum dot.

Universal quantum computation requires not only high-fidelity initialization, readout, and single-qubit
gates, but two-qubit entangling gates as well. While there has also been progress in achieving the
latter\cite{vanWeperenPRL2011,VeldhorstNature2016,NowackScience2011,ShulmanScience2012,NicholArXiv}, with
two-qubit gate fidelities as high as 90\% reported in recent work\cite{NicholArXiv}, fidelities have not
yet reached the thresholds necessary for quantum error correction schemes\cite{FowlerPRA2012}. This is
due in large part to decoherence caused by environmental noise. The two main sources of noise are magnetic
field noise, hereafter referred to simply as field noise, due to both nuclear spins in the host
semiconductor (Overhauser noise)\cite{DeSousaPRB2003} and fluctuations in the applied magnetic field,
and charge noise due to charge fluctuations on nearby impurities or on the electrostatic gates used to
confine electrons, leading to noise in the exchange coupling between the spins\cite{HuPRL2005}. Field noise
is especially large in GaAs; in fact, it is the dominant source of noise in this material and cannot be
eliminated as the only stable isotopes of Ga and As have nonzero nuclear spin.  However, its effects can
be considerably reduced by dynamical decoupling\cite{ViolaPRA1998,CywinskiPRB2008,BluhmPRL2010,MalinkowskiArXiv}
or Bayesian estimation of Hamiltonian parameters\cite{ShulmanNatCommun2014,SergeevichPRA2011}.  On the
other hand, field noise is less of a problem in Si; of the three stable isotopes, only ${}^{29}$Si
has a nonzero nuclear spin, and the concentration of this isotope can be greatly reduced via isotopic
purification\cite{WitzelPRL2010}.  However, charge noise continues to be a serious problem in Si, and
the field noise arising from fluctuations in the applied magnetic field remains an issue.

Developing a theoretical understanding of the effects of noise on entanglement creation in a system of two
spin qubits is therefore of great importance for future progress in building a semiconductor-based quantum
computer. However, a systematic investigation of the effects of both field and charge noise on our ability
to create entanglement has not been conducted previously. A number of works have introduced dynamical noise-suppression
techniques, such as the dynamical decoupling and Bayesian estimation methods mentioned above, as well as
dynamically corrected gates\cite{WangNatCommun2012,KestnerPRL2013,WangPRA2014,KhodjastehPRA2012,BarnesSciRep2015,UhrigPRL2007,WitzelPRL2007,LeePRL2008,YaoPRL2007}, which partially cancel out the effects of noise by applying carefully designed pulse sequences. In addition, several works have studied the decoherence of
an initially prepared entangled state of two or more qubits subject to various types of noise\cite{YuPRB2003,AnnPRB2007,DePRB2011,BragarPRB2015,SzankowskiQIP2015}.
There has also been some theoretical work on the dynamics of two coupled electron spins under a constant exchange
coupling and applied magnetic field gradient.  Two early treatments\cite{CoishPRB2005,KlauserPRB2006} mostly
focused on field noise, but included a limited discussion of charge noise as well. In more recent
works, we have studied the effects of noise on single-qubit coherence\cite{BarnesPRB2016} and on the state
preservation of two exchange-coupled qubits\cite{DasSarmaPRB2016}. However, none of these works provides a
comprehensive analysis of how noise limits entanglement generation in two-spin-qubit systems subject to realistic
noise. The question of how much noise is tolerable in the implementation of two-qubit gates at the error
correction threshold remains unknown. The goal of the present work is to address this question, which is
obviously of vital importance if semiconductor spin quantum computing is going to be a practical reality
in the future.

Most of the aforementioned works make use of the quasistatic bath approximation\cite{CywinskiAPP2011} in
which the noise is modeled by averaging the return probability over a Gaussian distribution of magnetic
fields and exchange couplings.  It is well known that the actual field and charge noise in spin qubit
experiments both have a complicated frequency dependence. Much work has been devoted to measuring this
dependence\cite{MedfordPRL2012,DialPRL2013} since it plays an important role in experiments that study or
manipulate spin evolution over time scales exceeding a few hundred nanoseconds. On the other hand, for
experiments that focus on shorter time scales, the Gaussian quasistatic model has been shown to work well
in fitting experimental data\cite{MartinsPRL2016,EngSciAdv2015,BarnesPRB2016,NederPRB2011}. Since our focus
here is on characterizing the effects of noise in these types of experiments, we employ the quasistatic
model throughout this work. While a detailed investigation of the effects of frequency-dependent noise is
important, it is beyond the scope of our present work.

The objective of our work is to calculate the fidelity of entanglement generation of two Heisenberg-coupled
electron spins in the presence of field and charge noise, starting from the ``classical'' unentangled
state $\ket{\uparrow\downarrow}$.  Suppose we allow the system to evolve from this initial state under the Heisenberg
Hamiltonian with no magnetic field gradient, i.e., under the influence of the exchange coupling $J$ alone.
In the complete absence of noise, the system will evolve into one of two maximally entangled states, $\ket{ME_1}=\frac{1}{\sqrt{2}}(\ket{\uparrow\downarrow}-i\ket{\downarrow\uparrow})$
and $\ket{ME_2}=\frac{1}{\sqrt{2}}(\ket{\uparrow\downarrow}+i\ket{\downarrow\uparrow})$, after times $t=\pi/2J$
and $t=3\pi/2J$, respectively.  The fidelities for producing these states are closely related to an intrinsic two-qubit
coherence time, which we denote by $T_2^{\ast}$, and to the steady-state return probability.  We emphasize that
this $T_2^{\ast}$, to be defined precisely in Sec.\ IIB below, is not the same as the free induction decay time
for a single electron spin---this is a two-electron property (the $2$ in the subscript of $T_2^{\ast}$ here refers
to two qubits or two coupled spins) that measures the suppression of entanglement creation between the two qubits.
For the ideal entangling operation, the return probability oscillates in time about $\tfrac{1}{2}$ with an amplitude of $\tfrac{1}{2}$.
Both the average (steady-state) value and the amplitude of oscillations are important for quantifying the maximum
amount of entanglement that can be generated. The intrinsic coherence time is a measure of the decay rate of the
oscillation amplitude of the return probability under the influence of disorder, and is thus a measure of the time
scale over which one can maximally entangle two electron spins. The steady-state value of the return probability
can also be influenced by noise, and deviations away from $\tfrac{1}{2}$ will lead to a further, time-independent
reduction of entanglement.  Our results can be used to determine to what extent the noise must be reduced in order
to achieve two-qubit entanglement fidelities that exceed quantum error correction thresholds.  We also calculate
the fidelities for producing the states, $\ket{ME_1}$ and $\ket{ME_2}$, which are defined as the disorder-averaged
probabilities that, after performing the operations that, ideally, would evolve the system from the unentangled
$\ket{\uparrow\downarrow}$ state to these maximally entangled states, we will measure the system to be in the appropriate
state\cite{NielsenBook,MagesanPRA2012}.

Our main findings are as follows.  We find that the coherence time decreases as we increase either type of noise,
but that charge noise has a noticeably greater effect.  As a very important check, we find that, in the absence
of magnetic disorder, the coherence time is inversely proportional to the charge noise strength for low disorder,
in agreement with the quasistatic bath approximation\cite{CywinskiAPP2011}; the fact that this is only approximate
in our case is due to the fact that we only allow for positive exchange coupling strengths, consistent with the
experimental situation in exchange-coupled semiconductor spin systems.  We also find that, in general, in the complete
absence of field noise, the steady-state return probability is always $\tfrac{1}{2}$.  This means that one can still
achieve maximal entanglement as long as the operation is done over a time scale much shorter than $T_2^{\ast}$.
If there is any field noise, however, then the steady-state return probability is larger than $\tfrac{1}{2}$, leading
to a reduction in the entanglement fidelity regardless of how quickly the entanglement is generated.  Our results
for the entangled state fidelities follow a somewhat different trend to that of $T_2^{\ast}$---we find that whereas
$T_2^{\ast}$ is much more sensitive to charge noise than field noise, the entanglement fidelity    for preparing
state $\ket{ME_1}$ is essentially equally sensitive to both types of noise. On the other hand, the fidelity for
preparing $\ket{ME_2}$ is affected less by field noise than by charge noise.  This follows from the basic observation
that it takes three times longer to prepare $\ket{ME_2}$ combined with the fact that charge noise tends to dominate
on longer time scales.  We note that similar analytic results to those that we find here are obtained for a triple
quantum dot system in Ref.\ \onlinecite{LaddPRB2012}.

Although these results would suggest that, in theory, charge noise would generally be a more severe problem for realizing
a maximally entangled state than field noise, we should note that, as mentioned before, field noise tends to be much larger
than charge noise in actual experiments, at least in GaAs, meaning that field noise is the dominant cause of decoherence
and thus loss of entanglement fidelity by the simple virtue of being the dominant source of noise in the system.  However,
it is possible to mitigate the effects of field noise by increasing the exchange coupling.  If we quantify the strengths of
the two types of noise present in the system by the standard deviations of their respective distributions, $\sigma_h$
for field noise and $\sigma_J$ for charge noise, then we will find that all of the quantities we calculate depend
on the noise strengths only in the dimensionless combinations, $\sigma_h/J_0$ and $\sigma_J/J_0$, where $J_0$ is the mean of the 
exchange-coupling distribution.  It has been shown that as $J_0$ is increased, $\sigma_h$ remains approximately constant, while $\sigma_J$ increases
roughly linearly.\cite{BarnesPRB2016}  Thus, even though field noise is strong in GaAs, it is possible to mitigate its effects
by simply increasing the exchange coupling, leaving charge noise as the main contributor to loss of entanglement fidelity.
This observation suggests that comparable entanglement fidelities can be achieved in both GaAs and Si systems, even
when isotopic purification is employed in the latter case, as one can reduce the effective dimensionless field noise
even in GaAs simply by increasing the exchange coupling strength.  Of course, single-qubit operations (i.e., spin rotations)
are also directly affected by field noise, but powerful dynamical decoupling techniques could mitigate field noise effects
on single qubit operations.  In the current work, we focus entirely on two-qubit exchange gate operations as the subject of
single-qubit operations for semiconductor spin qubit systems have already been studied rather exhaustively both theoretically
and experimentally with single qubit fidelity already reaching $99\%$ in some situations.

The rest of the paper is organized as follows.  In Sec.\ II, we review the Heisenberg Hamiltonian from which our
results ultimately derive, review the known results for the eigenstates, eigenenergies, and return probability, and
define the disorder average and intrinsic coherence time.  Section III is dedicated to our main results for the steady-state
return probability, coherence time, and fidelity.  We first give results in the limits of no field noise and
no charge noise, showing that we can obtain some purely analytical results in these limits.  We then give our exact
numerical results for the general case in which both types of noise are present.  We give our conclusions in Sec.\ IV.

\section{Model and definitions}
In this section, we review the Hamiltonian and eigenstates of the two-spin Heisenberg
model, as well as the return probability and magnetization for the system starting from a
``classical'' unentangled $\ket{\uparrow\downarrow}$ state.  We also give our definition
of the intrinsic coherence time, $T_2^{\ast}$, which we will be using throughout this paper.

\subsection{Heisenberg Hamiltonian}
Our Hamiltonian is that of two electron spins coupled via a disordered Heisenberg exchange
coupling subject to magnetic fields\cite{CoishPRB2005,KlauserPRB2006,DasSarmaPRB2016}:
\begin{equation}
H=J\vec{S}_1\cdot\vec{S}_2+h_1 S_{1,z}+h_2 S_{2,z}.
\end{equation}
Here, we take the magnetic fields $h_1$ and $h_2$ to have Gaussian distributions with a
mean of $h_0$ and standard deviation $\sigma_h$, hereafter called the ``strength'' of the
disorder,
\begin{equation}
f_{h_i}(h_i)=\frac{1}{\sigma_h\sqrt{2\pi}}e^{-(h_i-h_0)^2/2\sigma_h^2},
\end{equation}
and the exchange coupling to have a truncated Gaussian distribution (i.e., we restrict $J$ to
non-negative values) with a mean of $J_0$ and standard deviation $\sigma_J$,
\begin{equation}
f_J(J)=\frac{1}{\sigma_J\sqrt{2\pi}}\frac{2}{1+\mbox{erf}\left (\frac{J_0}{\sigma_J\sqrt{2}}\right )}e^{-(J-J_0)^2/2\sigma_J^2}.
\end{equation}
This Hamiltonian conserves the $z$ component of the total spin, $S_z=S_{1,z}+S_{2,z}$; we
will focus on the $S_z=0$ subspace from this point on, as there is only one eigenstate in
each of the $S_z=\pm 1$ subspaces, and therefore the dynamics of the system within these
subspaces will be trivial.

The effective Hamiltonian describing the $S_z=0$ subspace is
\begin{equation}
H_{\text{eff}}=\tfrac{1}{2}J\sigma_x+\tfrac{1}{2}\delta h\sigma_z-\tfrac{1}{4}J,
\end{equation}
where $\delta h=h_1-h_2$.  This Hamiltonian is written in the $\ket{\uparrow\downarrow}$
and $\ket{\downarrow\uparrow}$ basis, with the former being the ``top'' component of our
spinors.  Our subsequent calculations will be simplified by noting that the distribution
of $\delta h$ is also Gaussian, with zero mean and standard deviation $\sigma_h\sqrt{2}$:
\begin{equation}
f_{\delta h}(\delta h)=\frac{1}{2\sigma_h\sqrt{\pi}}e^{-(\delta h)^2/4\sigma_h^2}.
\end{equation}
Diagonalizing the effective Hamiltonian\cite{CoishPRB2005,KlauserPRB2006,DasSarmaPRB2016},
one finds that the energy eigenvalues are
\begin{equation}
E_{\pm}=-\tfrac{1}{4}J\pm\tfrac{1}{2}\sqrt{J^2+(\delta h)^2},
\end{equation}
and the associated eigenstates are
\begin{equation}
\psi_{\pm}=
\begin{bmatrix}
\frac{1}{\sqrt{2}}\sqrt{1\pm\frac{\delta h}{\sqrt{J^2+(\delta h)^2}}} \\
\pm\frac{1}{\sqrt{2}}\sqrt{1\mp\frac{\delta h}{\sqrt{J^2+(\delta h)^2}}}
\end{bmatrix}.
\end{equation}
Let us now consider initializing the system in the $\ket{\uparrow\downarrow}$ state.  We
now give the result for the return probability, $P_{\uparrow\downarrow}(t)=\left |\braket{\uparrow\downarrow|\Psi(t)}\right |^2$,
which is\cite{KlauserPRB2006,DasSarmaPRB2016}
\begin{equation}
P_{\uparrow\downarrow}(t)=1-\frac{J^2}{J^2+(\delta h)^2}\sin^2\left [\tfrac{1}{2}\sqrt{J^2+(\delta h)^2}t\right ]. \label{Eq:PUpDownSR}
\end{equation}
We denote the disorder average of a quantity $A$ as $[A]_{\alpha}$, which is defined simply as
\begin{equation}
[A]_{\alpha}=\int_{-\infty}^{\infty}d(\delta h)\,\int_{0}^{\infty}dJ\,f_{\delta h}(\delta h)f_J(J)A. \label{Eq:DisorderAvg}
\end{equation}
In general, this average must be evaluated numerically; however, as we will see below, there
are special cases in which it is possible to obtain analytical results.

\subsection{Definition of $T_2^{\ast}$}
We now define the intrinsic coherence time, $T_2^{\ast}$.  We will see below that the disorder-averaged
return probability shows oscillations that decay in amplitude, tending toward a steady-state value.  We
define $T_2^{\ast}$ as the time that it takes for the amplitude of these oscillations about the steady-state
value to decay to $1/e$ times the initial amplitude.  We wish to emphasize two points about $T_2^{\ast}$.
First, this definition is purely operational---it is independent of the detailed time dependence of
the amplitude and is used purely for convenience.  Any other reasonable definition of $T_2^{\ast}$ will
result in the same conclusions.  Second, $T_2^{\ast}$ here is not related to the free-induction decay time
of a single qubit---rather, it is a two-qubit property, as already emphasized in Sec.\ I.  It is
a measure of the rate at which our ability to entangle the two qubits is suppressed as the time scale
of entanglement generation is increased. Also note that the dimensionless number $J_0 T_2^{\ast}$ gives
the number of coherent oscillations exhibited in the return probability before it decays.

We may connect this decay of the oscillation amplitude of the return probability to entanglement of the
two electron spins as follows.  Let us first consider the case with no noise and no magnetic field gradient.
In this case, we find that the return probability is just the oscillatory function,
\begin{equation}
P_{\uparrow\downarrow}(t)=\frac{1+\cos{Jt}}{2}. \label{Eq:PUpDownSRNoMagGrad}
\end{equation}
We therefore see that the return probability oscillates around $\tfrac{1}{2}$ with an amplitude of
$\tfrac{1}{2}$.  If we let the system evolve for a time $t=\pi/2J$, then the return probability will
be exactly $\tfrac{1}{2}$, while the state of the system will be the maximally entangled state,
\begin{equation}
e^{-i\pi\sigma_x/4}\ket{\uparrow\downarrow}=\frac{1}{\sqrt{2}}(\ket{\uparrow\downarrow}-i\ket{\downarrow\uparrow})=\ket{ME_1},
\end{equation}
which differs from the singlet and triplet states by only single-qubit operations.  We may also realize
a maximally entangled state by evolving the system for a time $t=3\pi/2J$; in this case, we obtain
\begin{equation}
e^{-3i\pi\sigma_x/4}\ket{\uparrow\downarrow}=-\frac{1}{\sqrt{2}}(\ket{\uparrow\downarrow}+i\ket{\downarrow\uparrow})=-\ket{ME_2}.
\end{equation}
We thus see that a return probability of the form given in Eq.\ \eqref{Eq:PUpDownSRNoMagGrad} indicates
an operation that is capable of producing a maximally entangled state. These operations are in fact $\sqrt{\hbox{SWAP}}$
gates up to single-qubit rotations. Any deviations from this form, whether due to the amplitude decaying
or due to a shift in the value about which the probability oscillates, indicate a reduction in our ability
to produce such maximal entanglement.  Therefore, one may view $T_2^{\ast}$ as an operationally defined
characteristic time scale over which the two-qubit entanglement decays.

\section{Return probability and coherence time}
We now present our results for the disorder-averaged return probability $[P_{\uparrow\downarrow}(t)]_{\alpha}$
and the intrinsic coherence times $T_2^{\ast}$ extracted from it.  We first present the general
formula for the return probability, and then show that, in the $\sigma_h=0$ and $\sigma_J=0$ limits,
we can obtain some closed-form analytical results.

If we substitute Eq.\ \eqref{Eq:PUpDownSR} into Eq.\ \eqref{Eq:DisorderAvg}, use the trigonometric
identity, $\sin^2{\theta}=\frac{1-\cos{2\theta}}{2}$, and rewrite in terms of the real part of a
complex-valued expression, we obtain
\begin{widetext}
\begin{eqnarray}
[P_{\uparrow\downarrow}(t)]_{\alpha}&=&1-\frac{1}{2\pi\sigma_h\sigma_J\sqrt{2}\left [1+\mbox{erf}\left (\frac{J_0}{\sigma_J\sqrt{2}}\right )\right ]}\int_{-\infty}^{\infty}d(\delta h)\,\int_0^{\infty}dJ\,\frac{J^2}{J^2+(\delta h)^2}e^{-(\delta h)^2/4\sigma_h^2}e^{-(J-J_0)^2/2\sigma_J^2} \cr
&+&\frac{1}{2\pi\sigma_h\sigma_J\sqrt{2}\left [1+\mbox{erf}\left (\frac{J_0}{\sigma_J\sqrt{2}}\right )\right ]}\mbox{Re}\left [\int_{-\infty}^{\infty}d(\delta h)\,\int_0^{\infty}dJ\,\frac{J^2}{J^2+(\delta h)^2}e^{-(\delta h)^2/4\sigma_h^2}e^{-(J-J_0)^2/2\sigma_J^2}e^{i\sqrt{J^2+(\delta h)^2}t}\right ]. \nonumber \\
\end{eqnarray}
\end{widetext}
We now show that the first two terms of the above expression give the steady-state return probability,
which we will denote by $P_S$, so that all of the oscillations about said steady-state value come from
the third term.  Let us denote the integral in the third term by $I(t)$:
\begin{eqnarray}
I(t)&=&\int_{-\infty}^{\infty}d(\delta h)\,\int_0^{\infty}dJ\,\frac{J^2}{J^2+(\delta h)^2}e^{-(\delta h)^2/4\sigma_h^2} \cr
&\times&e^{-(J-J_0)^2/2\sigma_J^2}e^{i\sqrt{J^2+(\delta h)^2}t}
\end{eqnarray}
If we rewrite this in polar coordinates, $\delta h=r\cos{\theta}$ and $J=r\sin{\theta}$, we obtain
\begin{eqnarray}
I(t)&=&2\int_{0}^{\infty}dr\,\int_0^{\pi/2}d\theta\,r\sin^2{\theta}e^{-r^2\cos^2{\theta}/4\sigma_h^2} \cr
&\times&e^{-(r\sin{\theta}-J_0)^2/2\sigma_J^2}e^{irt}.
\end{eqnarray}
We now take the Fourier transform,
\begin{equation}
I(\omega)=\int_{-\infty}^{\infty}dt\,I(t)e^{-i\omega t},
\end{equation}
of this expression, obtaining, for $\omega\geq 0$,
\begin{eqnarray}
I(\omega)&=&2\int_{0}^{\infty}dr\,\int_0^{\pi/2}d\theta\,r\sin^2{\theta}e^{-r^2\cos^2{\theta}/4\sigma_h^2} \cr
&\times&e^{-(r\sin{\theta}-J_0)^2/2\sigma_J^2}\cdot 2\pi\delta(\omega-r) \cr
&=&4\pi\omega\int_0^{\pi/2}d\theta\,\sin^2{\theta}e^{-\omega^2\cos^2{\theta}/4\sigma_h^2}e^{-(\omega\sin{\theta}-J_0)^2/2\sigma_J^2}. \nonumber \\
\end{eqnarray}
We thus see that $I(\omega=0)=0$, and thus the third term in $[P_{\uparrow\downarrow}(t)]_{\alpha}$
simply represents oscillations about the steady-state value, which is given by the first two terms:
\begin{eqnarray}
P_S&=&1-\frac{1}{2\pi\sigma_h\sigma_J\sqrt{2}\left [1+\mbox{erf}\left (\frac{J_0}{\sigma_J\sqrt{2}}\right )\right ]} \cr
&\times&\int_{-\infty}^{\infty}d(\delta h)\,\int_0^{\infty}dJ\,\frac{J^2}{J^2+(\delta h)^2}e^{-(\delta h)^2/4\sigma_h^2} \cr
&\times&e^{-(J-J_0)^2/2\sigma_J^2}.
\end{eqnarray}

\subsection{$\sigma_h=0$ limit}
Now we turn our attention to the $\sigma_h=0$ limit.  In this case, the steady-state return probability
is just
\begin{eqnarray}
P_S&=&1-\frac{1}{\sigma_J\sqrt{2\pi}\left [1+\mbox{erf}\left (\frac{J_0}{\sigma_J\sqrt{2}}\right )\right ]}\int_0^{\infty}dJ\,e^{-(J-J_0)^2/2\sigma_J^2} \cr
&=&\frac{1}{2}.
\end{eqnarray}

We can also obtain a closed-form analytical solution for the return probability as a function of time.  Our
formula for $[P_{\uparrow\downarrow}(t)]_{\alpha}$ becomes
\begin{eqnarray}
[P_{\uparrow\downarrow}(t)]_{\alpha}&=&\frac{1}{2}+\frac{1}{\sigma_J\sqrt{2\pi}\left [1+\mbox{erf}\left (\frac{J_0}{\sigma_J\sqrt{2}}\right )\right ]} \cr
&\times&\mbox{Re}\left [\int_0^{\infty}dJ\,e^{-(J-J_0)^2/2\sigma_J^2}e^{iJt}\right ].
\end{eqnarray}
The integral can be evaluated analytically in terms of the error function; we obtain
\begin{eqnarray}
[P_{\uparrow\downarrow}(t)]_{\alpha}&=&\frac{1}{2}+\frac{1}{2\left [1+\mbox{erf}\left (\frac{J_0}{\sigma_J\sqrt{2}}\right )\right ]}e^{-\sigma_J^2 t^2/2} \cr
&\times&\left\{\cos{J_0 t}+\mbox{Re}\left [e^{iJ_0 t}\mbox{erf}\left (\frac{J_0}{\sigma_J\sqrt{2}}+i\frac{\sigma_J t}{\sqrt{2}}\right )\right ]\right\}. \nonumber \\
\end{eqnarray}

We now show that, in the limit, $\sigma_J\ll J_0$, it is possible to obtain an approximate analytical expression
for $T_2^{\ast}$.  In this limit, we may drop the imaginary part of the argument of the error function, obtaining
\begin{equation}
[P_{\uparrow\downarrow}(t)]_{\alpha}\approx\tfrac{1}{2}(1+e^{-\sigma_J^2 t^2/2}\cos{J_0 t}).
\end{equation}
This approximation only holds if $t\ll\frac{J_0}{\sigma_J^2}$.  We see that, in this limit, the return probability
has a Gaussian decay towards its steady-state value.  We can now simply read off the value of $T_2^{\ast}$,
obtaining
\begin{equation}
T_2^{\ast}\approx\frac{\sqrt{2}}{\sigma_J}. \label{Eq:T2S_noSigh_SmallSigJ}
\end{equation}
For small $\sigma_J$ ($\ll J_0$), we thus see that our approximation is justified, though it will begin to break down for
larger $\sigma_J$.  We have thus recovered the result for $T_2^{\ast}$ found in the quasistatic bath approximation\cite{CywinskiAPP2011}.
Note that, unlike in the treatment of Ref.\ \onlinecite{CywinskiAPP2011}, we truncate the Gaussian distribution
to positive values of the exchange coupling only.  If we had not done so, then the above formula would in fact
be exact.

We should note that, because the return probability oscillates around $\tfrac{1}{2}$, the amplitude of the
oscillations must initially be $\tfrac{1}{2}$ because the probability at $t=0$ is $1$.  This indicates that
the decoherence caused by charge noise is due entirely to decay of the amplitude of these oscillations.

\subsection{$\sigma_J=0$ limit}
\begin{figure}[ht]
\includegraphics[width=0.5\columnwidth]{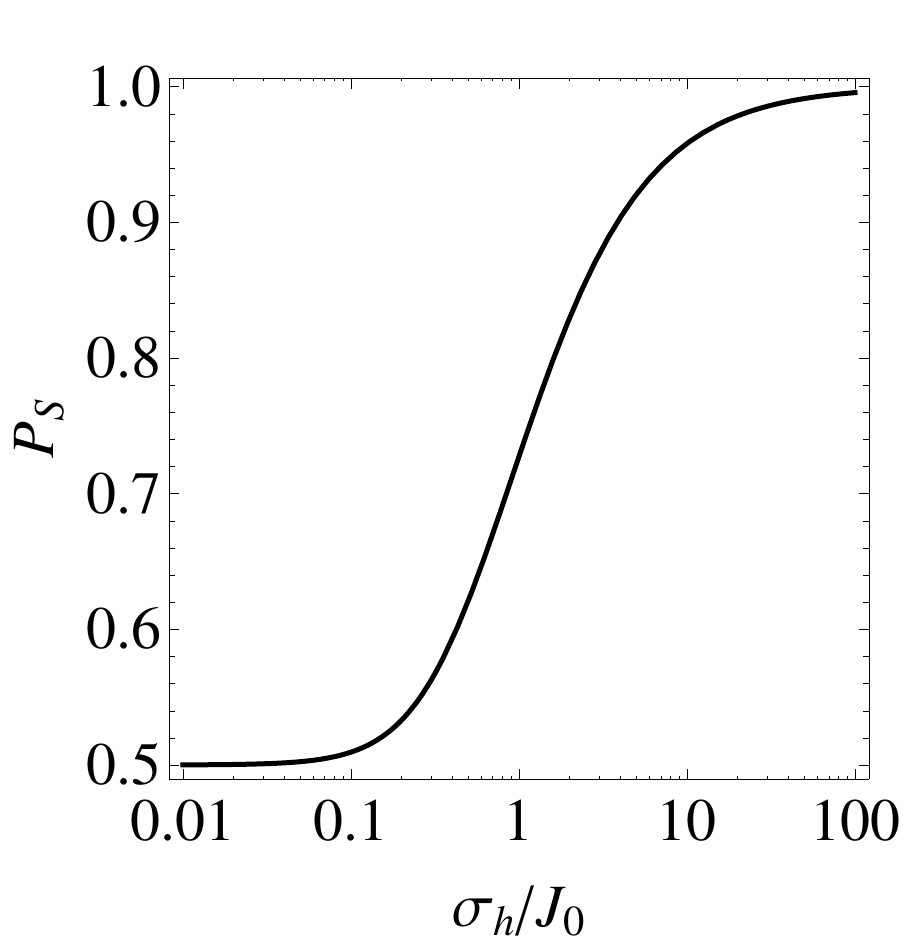}
\caption{Plot of $P_S$ as a function of $\sigma_h/J_0$ for $\sigma_J=0$.}\label{Fig:PS_FuncSigh_noSigJ}
\end{figure}

We now consider the $\sigma_J=0$ limit.  Here, the steady-state return probability becomes
\begin{eqnarray}
P_S&=&1-\frac{1}{4\sigma_h\sqrt{\pi}}\int_{-\infty}^{\infty}d(\delta h)\,\frac{J_0^2}{J_0^2+(\delta h)^2}e^{-(\delta h)^2/4\sigma_h^2} \cr
&=&1-\frac{J_0\sqrt{\pi}}{4\sigma_h}e^{J_0^2/4\sigma_h^2}\mbox{erfc}\left (\frac{J_0}{2\sigma_h}\right ).
\end{eqnarray}
We present a plot of this result in Fig.\ \ref{Fig:PS_FuncSigh_noSigJ}.  We thus see the ``memory retention''
effect pointed out in Ref.\ \onlinecite{DasSarmaPRB2016}.  However, this effect is
actually detrimental to our ability to realize a maximally entangled state.  This result necessarily implies
that the amplitude of the oscillations in the return probability will always be less than $\tfrac{1}{2}$.
This indicates that, in addition to the decay of the oscillations of the return probability, field noise
also causes decoherence by shifting the steady-state return probability to a value greater than $\tfrac{1}{2}$.
This dichotomy between quantum memory and quantum entanglement is understandable since any retained memory of
the initial non-entangled state can only hinder achieving maximal entanglement in the final state.

The expression for the return probability as a function of time, which cannot be evaluated in closed form, is
\begin{eqnarray}
[P_{\uparrow\downarrow}(t)]_{\alpha}&=&1-\frac{J_0\sqrt{\pi}}{4\sigma_h}e^{J_0^2/4\sigma_h^2}\mbox{erfc}\left (\frac{J_0}{2\sigma_h}\right ) \cr
&+&\frac{1}{4\sigma_h\sqrt{\pi}}\mbox{Re}\left [\int_{-\infty}^{\infty}d(\delta h)\,\frac{J_0^2}{J_0^2+(\delta h)^2}e^{-(\delta h)^2/4\sigma_h^2}\right. \cr
&\times&\left.e^{i\sqrt{J_0^2+(\delta h)^2}t}\right ].
\end{eqnarray}

\subsection{Numerical results for $\sigma_h=0$ and for $\sigma_J=0$}
\begin{figure}[ht]
\includegraphics[width=0.49\columnwidth]{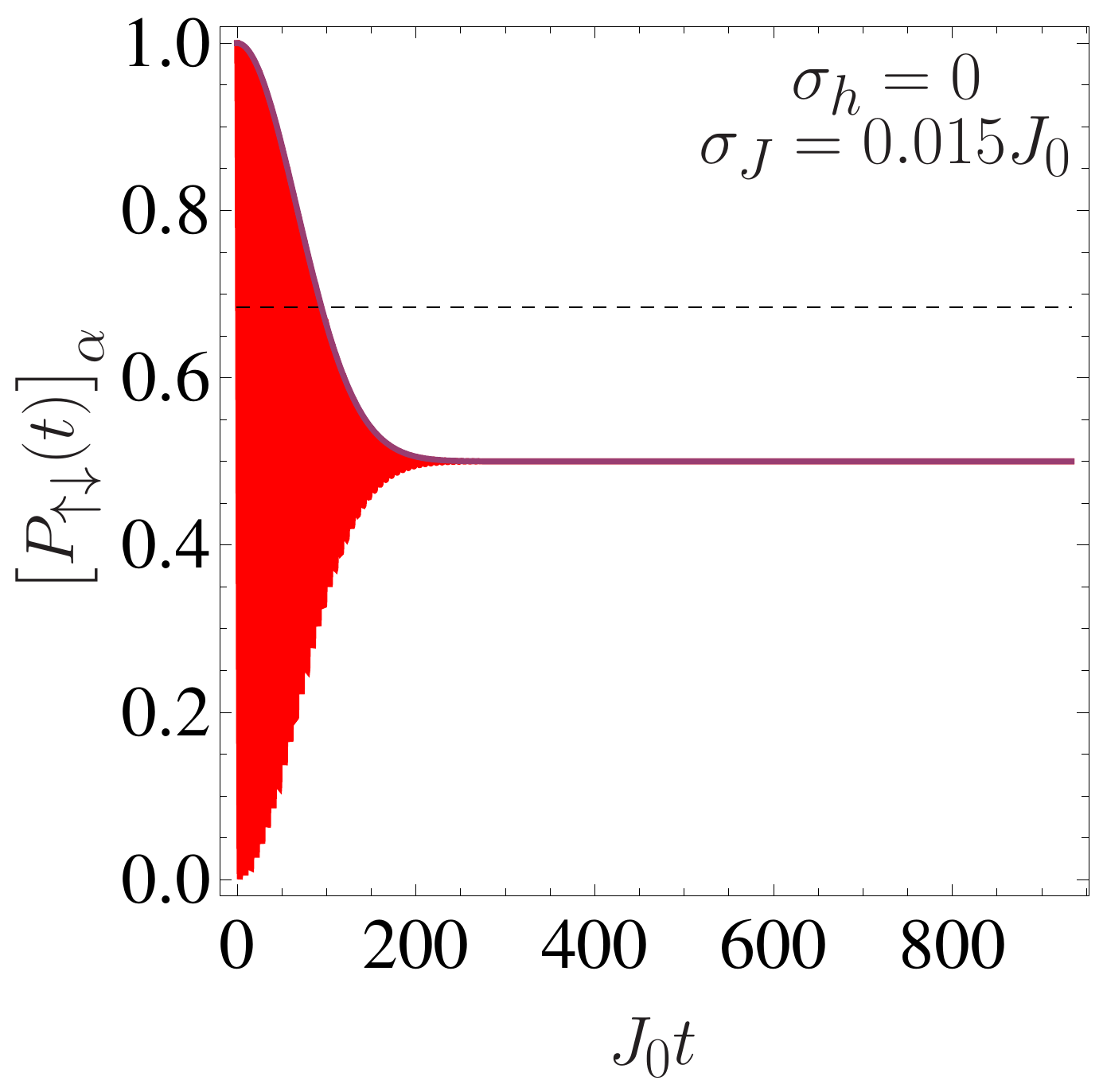}
\includegraphics[width=0.49\columnwidth]{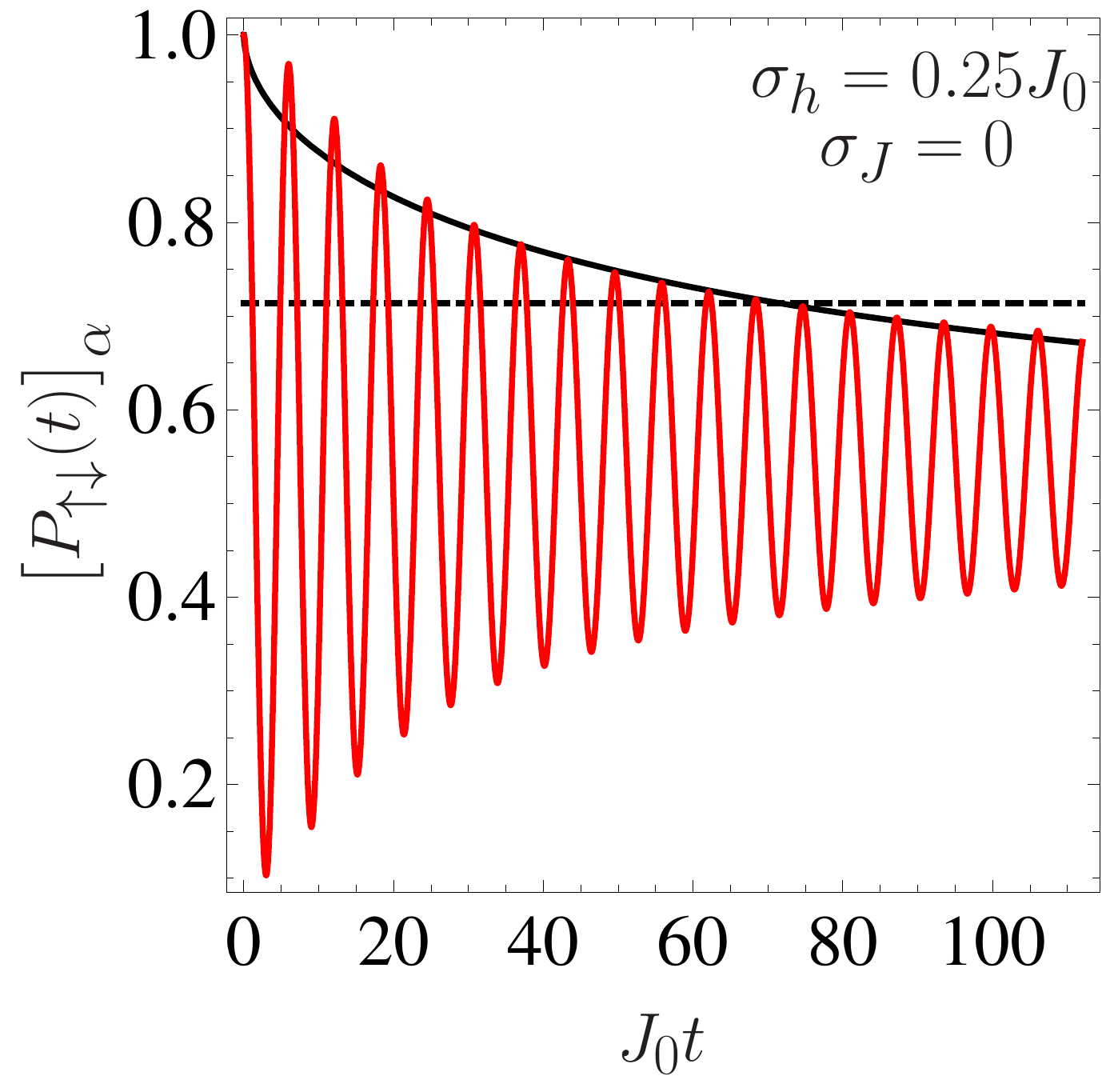}
\includegraphics[width=0.49\columnwidth]{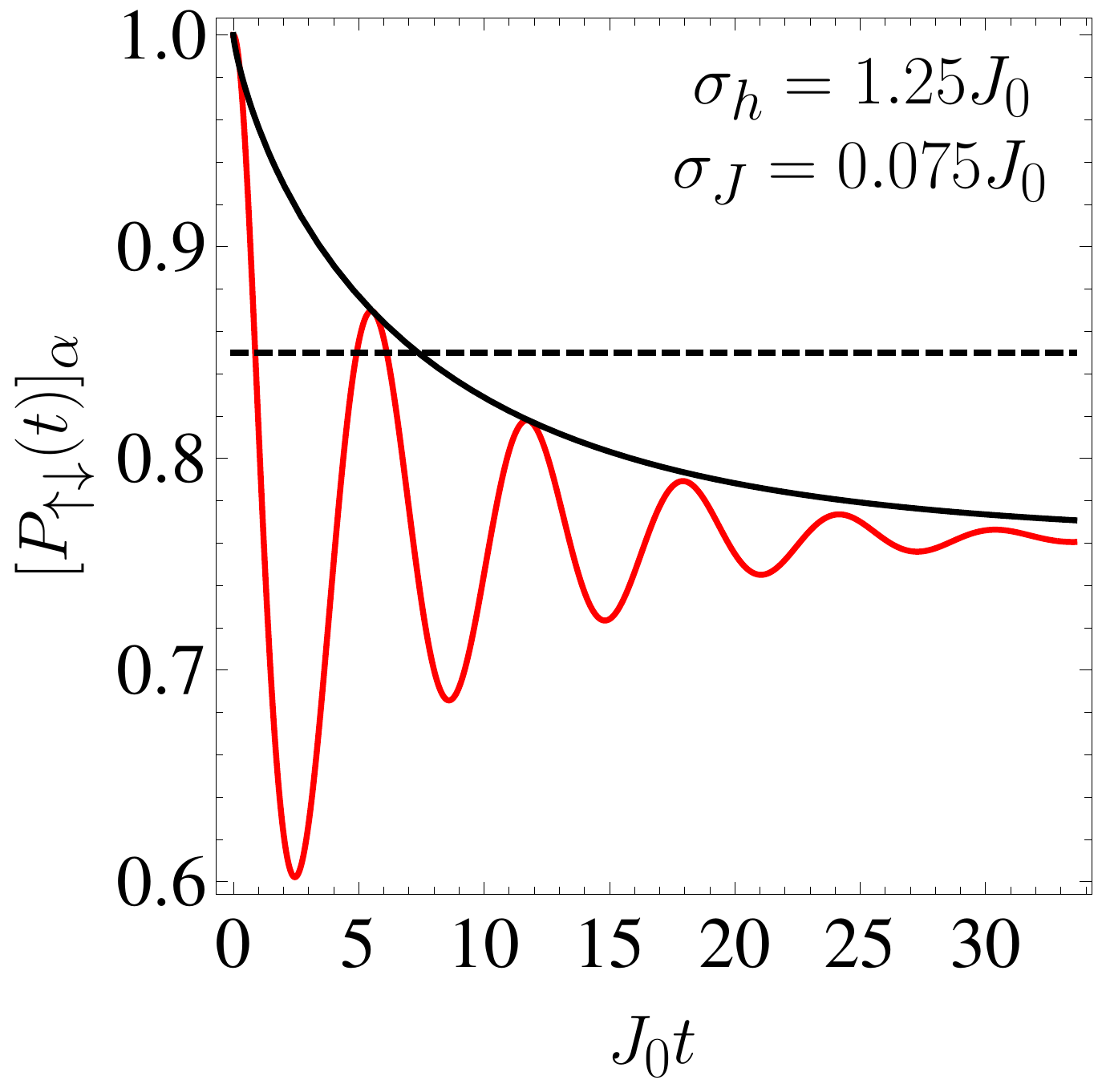}
\caption{Examples of our envelope-fitting procedure for $\sigma_h=0$ and $\sigma_J/J_0=0.015$ (top left),
$\sigma_h/J_0=0.25$ and $\sigma_J=0$ (top right), and $\sigma_h/J_0=1.25$ and $\sigma_J/J_0=0.075$ (bottom).
Here, the red curve is the return probability as a function of time, the black curve is the approximation
to the envelope given in Eq.\ \eqref{Eq:ApproxEnv}, and the dashed line is a guide showing where the amplitude
of the oscillations in the return probability becomes $1/e$ times its maximum value.  In these examples, the
values of $\alpha$ and $T_2^{\ast}$ are $\alpha=2$ and $J_0 T_2^{\ast}=94.5$ (top left), $\alpha=0.575$
and $J_0 T_2^{\ast}=71.5$ (top right), and $\alpha=0.8$ and $J_0 T_2^{\ast}=7.37$ (bottom).}
\label{Fig:T2S_Samples}
\end{figure}

We now present our numerical results in the above two limits.  In general, we cannot determine $T_2^{\ast}$
from an analytical formula, and thus we must extract it numerically from the return probability curve.  We
now describe how we do so.  We attempt to find a curve of the form,
\begin{equation}
P_E(t)=P_S+(1-P_S)e^{-(t/T_2^{\ast})^\alpha}, \label{Eq:ApproxEnv}
\end{equation}
which closely approximates the ``envelope'' of the return probability curve in the vicinity of the time at which
the amplitude of the return probability appears to reach $1/e$ times its value at $t=0$.  We simply adjust
$T_2^{\ast}$ and $\alpha$ until we obtain the most satisfactory fit, and then we read off $T_2^{\ast}$.  We
show several illustrations of this fit in Fig.\ \ref{Fig:T2S_Samples} (the third example given is for a case
in which neither $\sigma_h$ nor $\sigma_J$ are zero; we treat this general case in Sec.\ III D).  We
emphasize that we do not attempt to fit the exact ``envelope'' of the return probability curve for all times,
but only in the vicinity of $T_2^{\ast}$.  Our typical best fit value of $\alpha$ falls in the range of $0.5$--$2.0$
depending on the details of the parameters (i.e., $\sigma_h$, $\sigma_J$, $J_0$, etc.) although no significance should
be attached to the precise value of $\alpha$ since all we are trying to do here is to extract an operationally
meaningful value of $T_2^{\ast}$ through an accurate numerical fitting to the exact two-qubit dynamics.

\begin{figure}[ht]
\includegraphics[width=0.49\columnwidth]{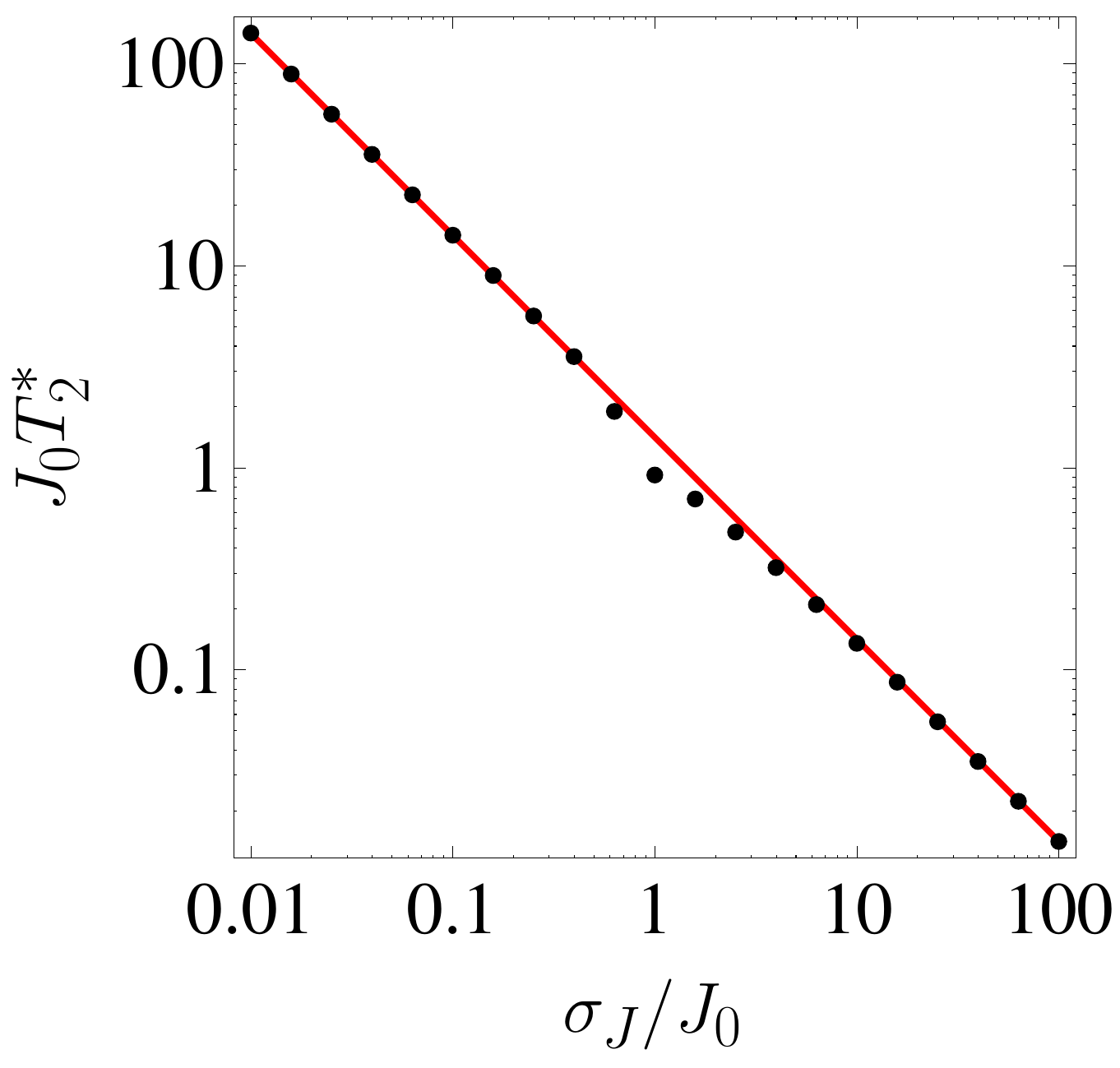}
\includegraphics[width=0.49\columnwidth]{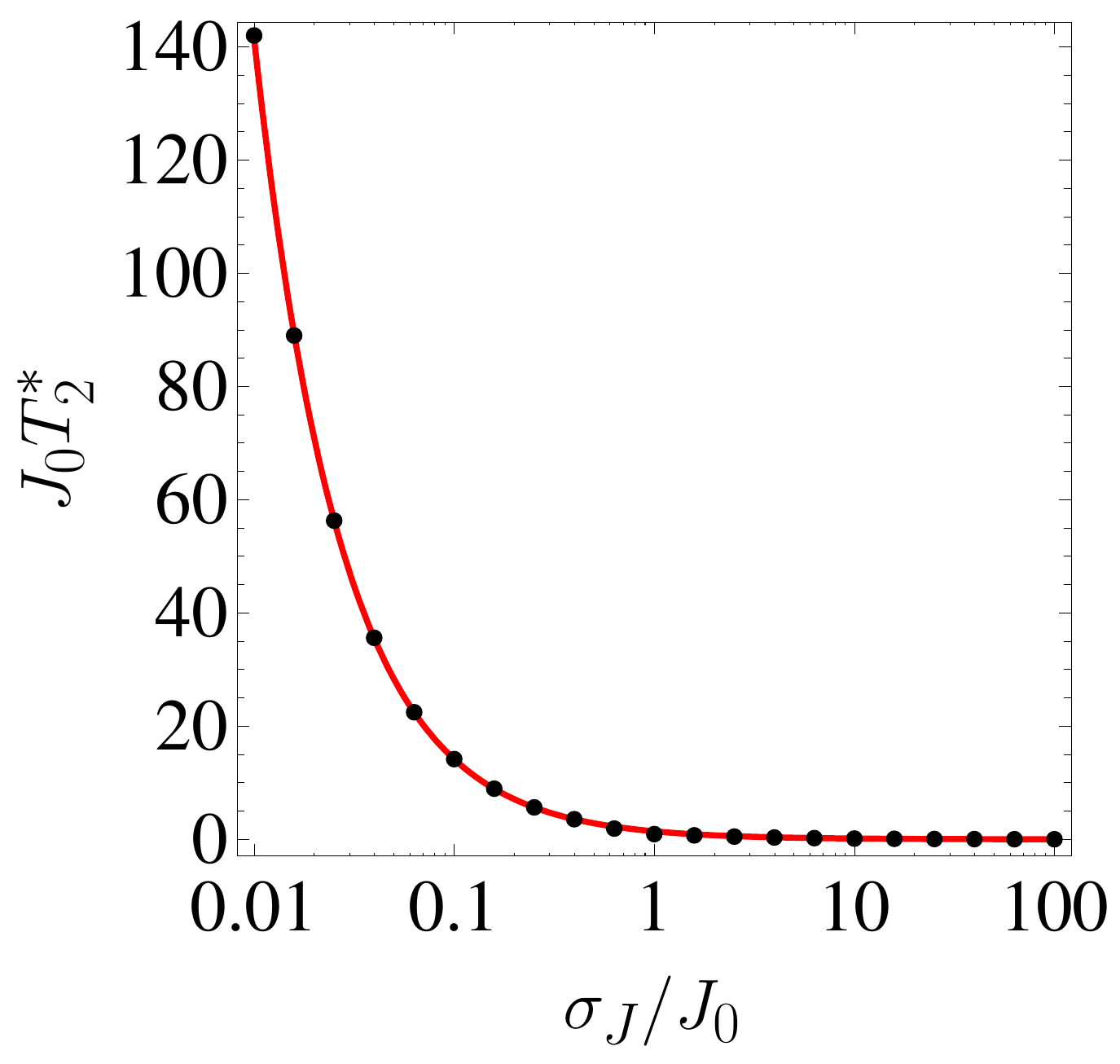}
\caption{Plot of $J_0 T_2^{\ast}$ for $\sigma_h=0$ as a function of $\sigma_J/J_0$ on a log-log scale (left)
and a log-linear scale (right).  The black dots are the numerical values extracted from plots of the return
probability as a function of time, while the red curves are the analytic approximation, Eq.\ \eqref{Eq:T2S_noSigh_SmallSigJ},
derived for small values of $\sigma_J/J_0$.}
\label{Fig:T2S_noSigh}
\end{figure}

\begin{figure}[ht]
\includegraphics[width=0.49\columnwidth]{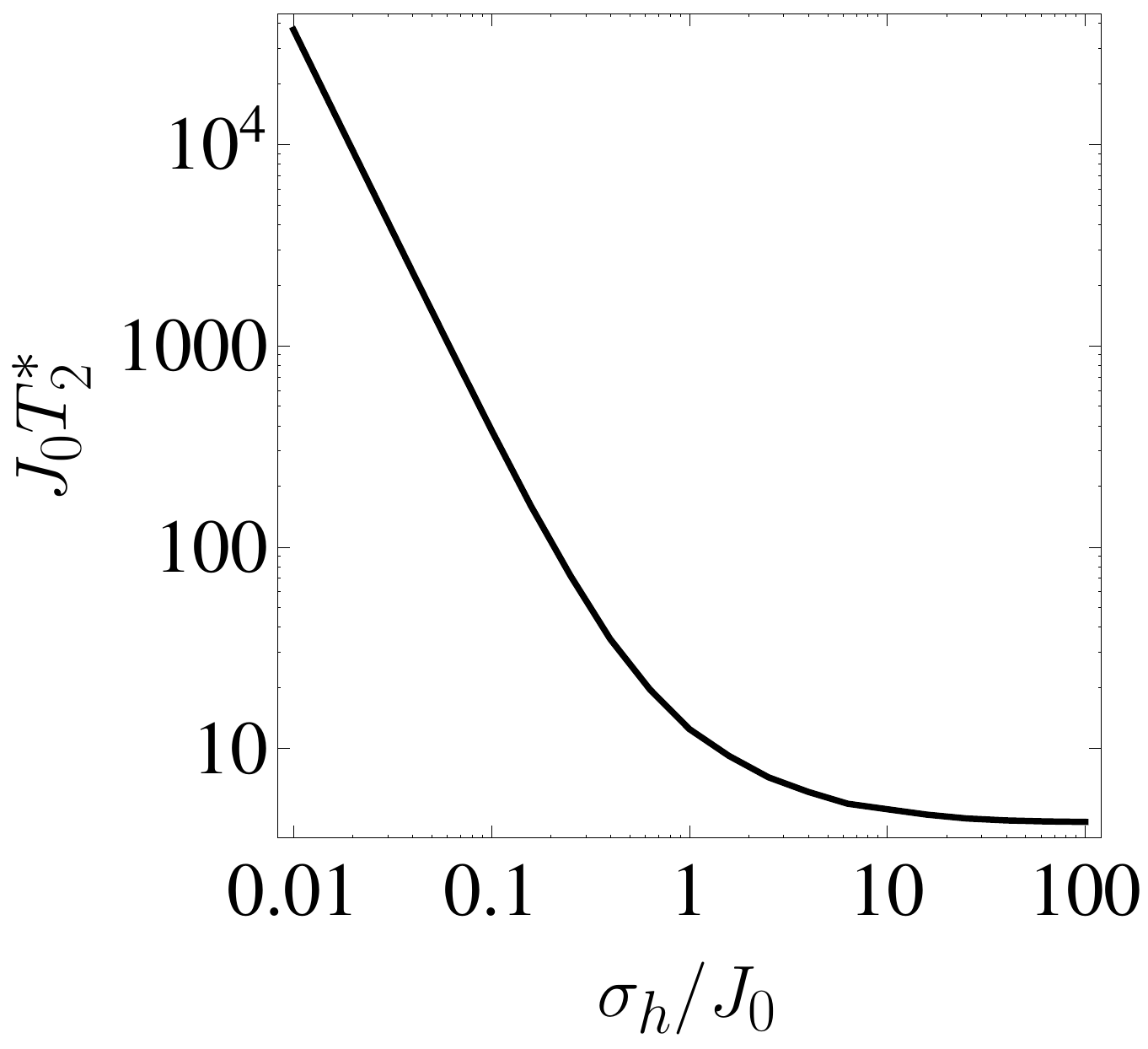}
\includegraphics[width=0.49\columnwidth]{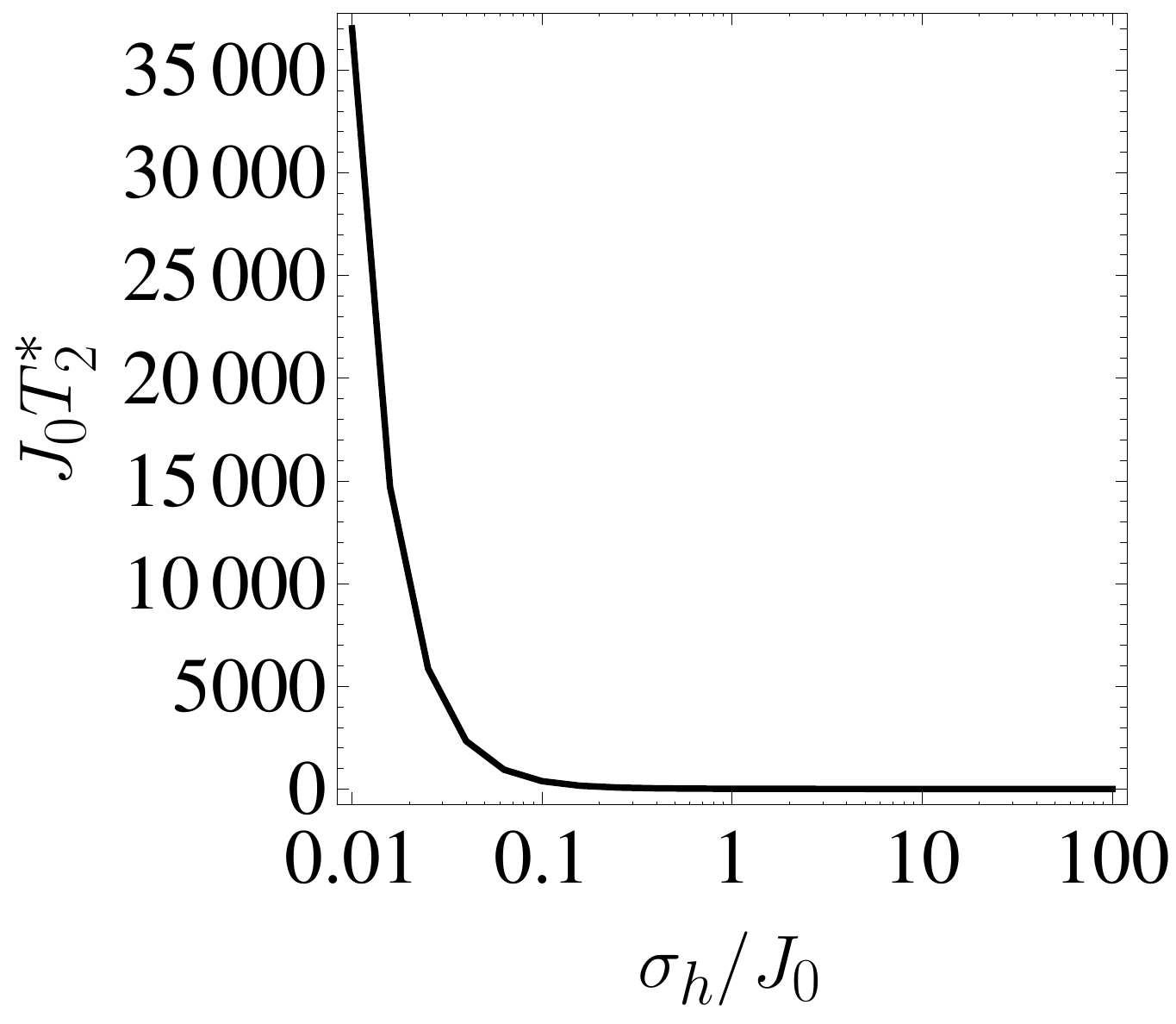}
\caption{Plot of $J_0 T_2^{\ast}$ for $\sigma_J=0$ as a function of $\sigma_h/J_0$ on a log-log scale (left)
and a log-linear scale (right).}
\label{Fig:T2S_noSigJ}
\end{figure}

We now give our results so obtained in the $\sigma_h=0$ and $\sigma_J=0$ limits.  We show our results for
the $\sigma_h=0$ limit in Fig.\ \ref{Fig:T2S_noSigh} and those for the $\sigma_J=0$ limit in Fig.\ \ref{Fig:T2S_noSigJ}.

There are two observations we make about these results.  First, we have verified our approximation
for $T_2^{\ast}$ for $\sigma_h=0$ and small $\sigma_J$, but also see that the approximation appears to work very
well, not just when $\sigma_J$ is comparable to $J_0$, but even when it is large.  The second observation is
that the $T_2^{\ast}$ values that we obtain when $\sigma_h=0$ and $\sigma_J$ varies are much smaller than those
that we obtain for $\sigma_J=0$ for comparable values of $\sigma_h$.  We will see later that this results in
a given amount of field noise causing less of an overall loss of entanglement fidelity than an identical amount
of charge noise would, despite the fact that field noise causes both a decay of the oscillations in the return
probability and a shift of the steady-state return probability, while charge noise only causes a decay of oscillations.

\subsection{General results}
\begin{figure}[ht]
\includegraphics[width=0.49\columnwidth]{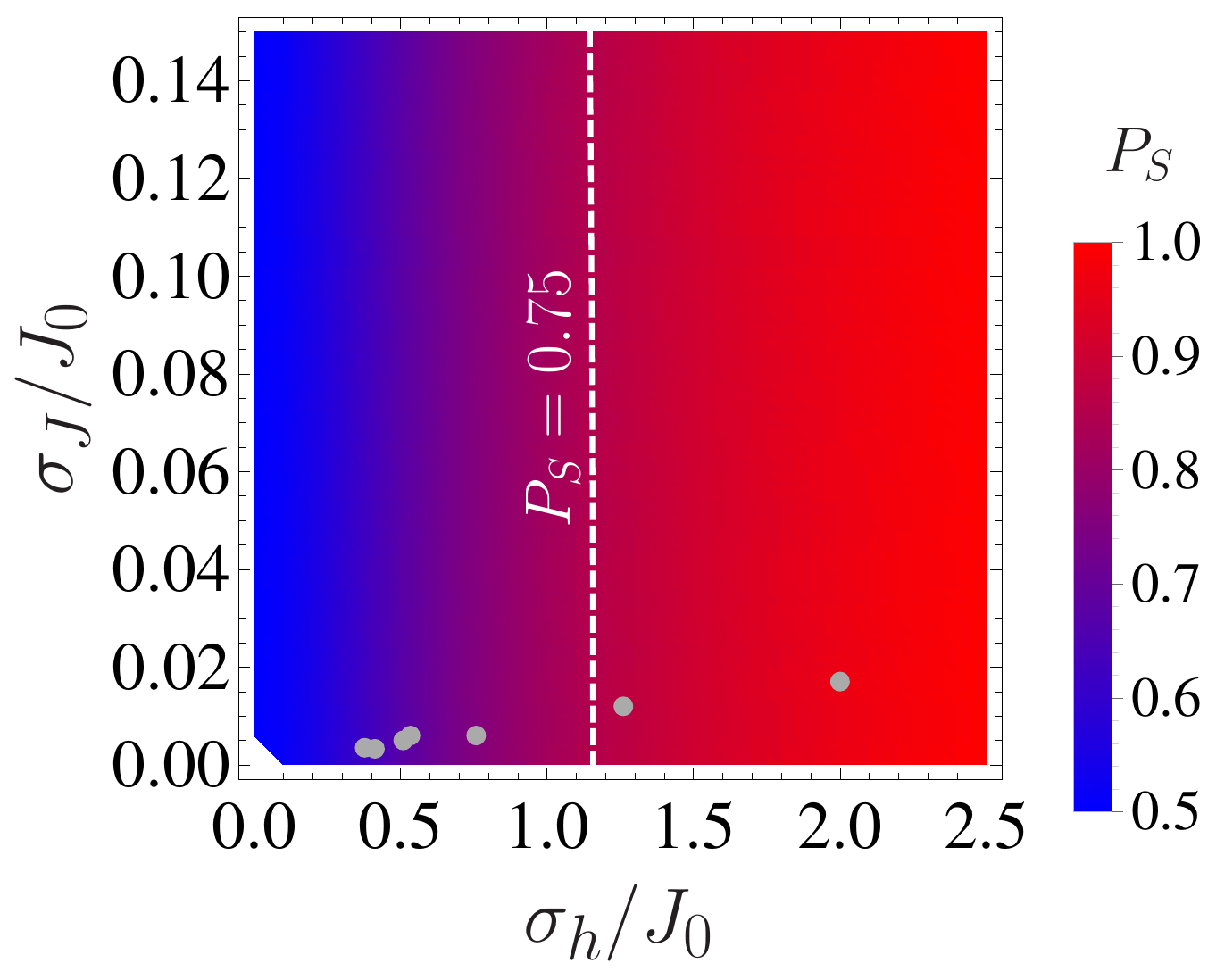}
\includegraphics[width=0.49\columnwidth]{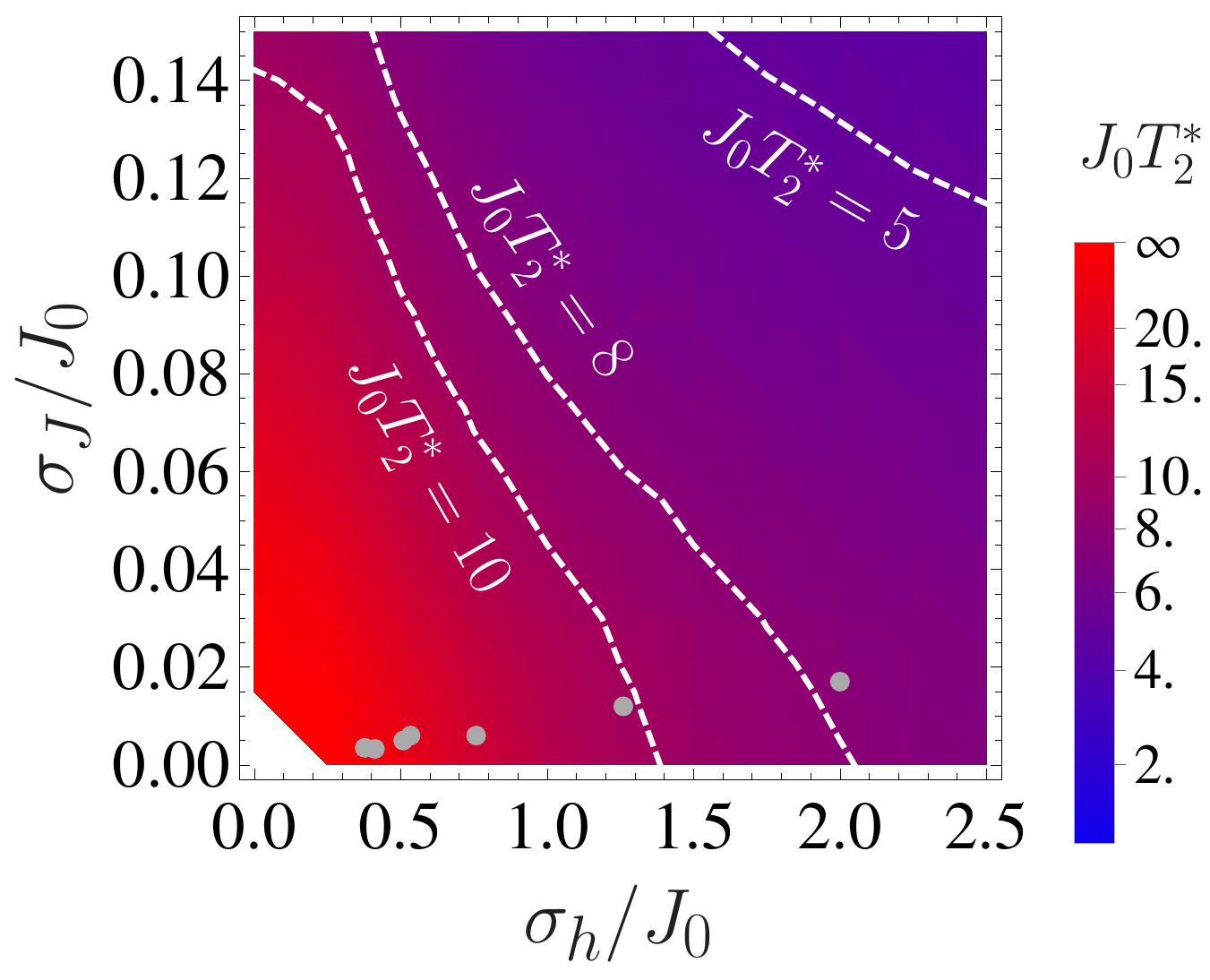}
\caption{(Left) Plot of the steady-state return probability $P_S$ as a function of $\sigma_h/J_0$ and
$\sigma_J/J_0$ over the region, $0\leq\sigma_h/J_0\leq 2.5$ and $0\leq\sigma_J/J_0\leq 0.15$.  The dashed
line indicates points at which $P_S=0.75$. (Right) Plot of the (dimensionless) coherence time, $J_0 T_2^{\ast}$,
over the same domain.  The dashed lines indicate points at which $J_0 T_2^{\ast}=5$, $8$, and $10$.  In
both plots, the gray dots represent the strength of the noise present in the experiments described in
Ref.\ \onlinecite{MartinsPRL2016}.}
\label{Fig:Results_10_100}
\end{figure}

\begin{figure}[ht]
\includegraphics[width=0.49\columnwidth]{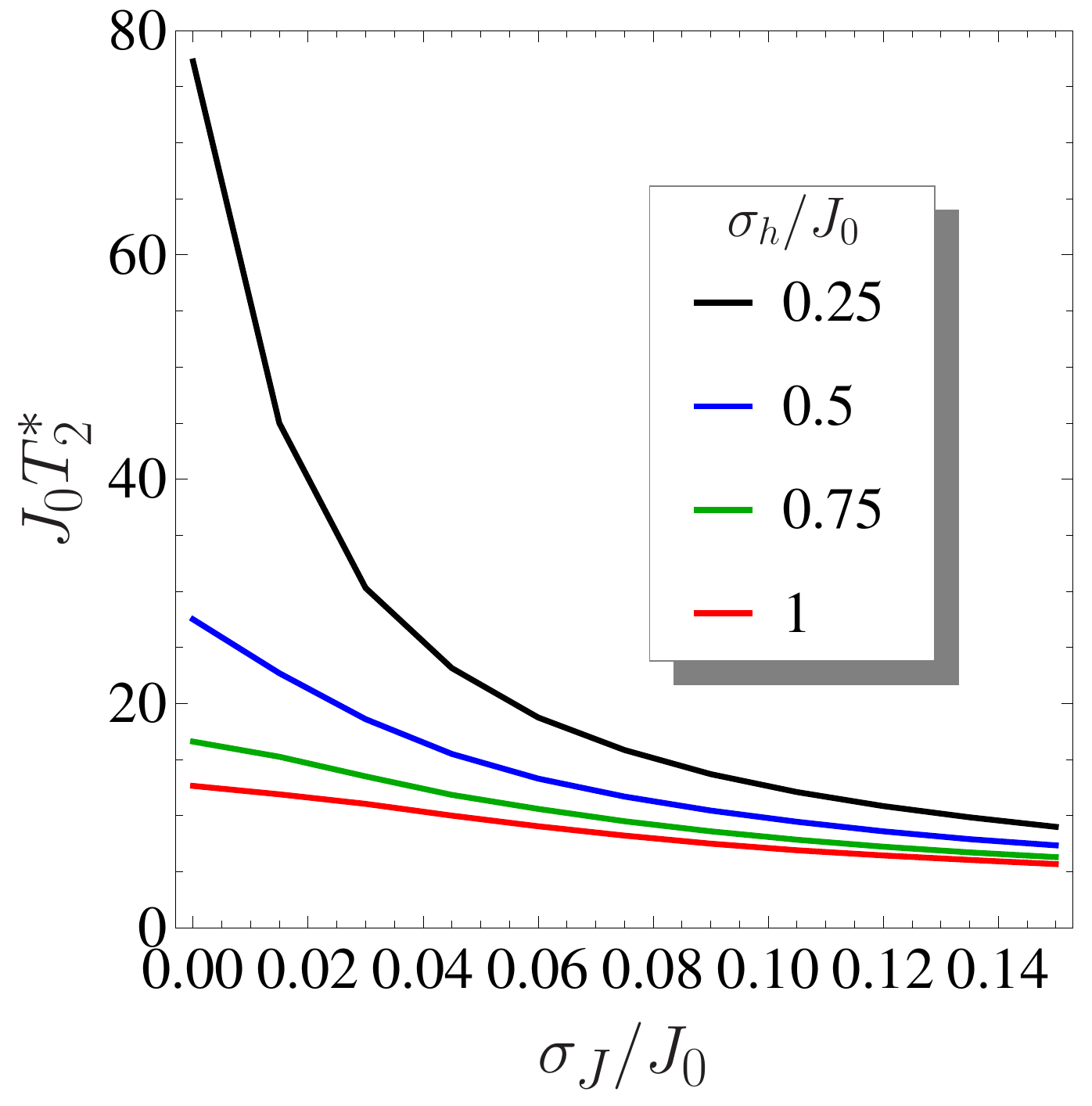}
\includegraphics[width=0.49\columnwidth]{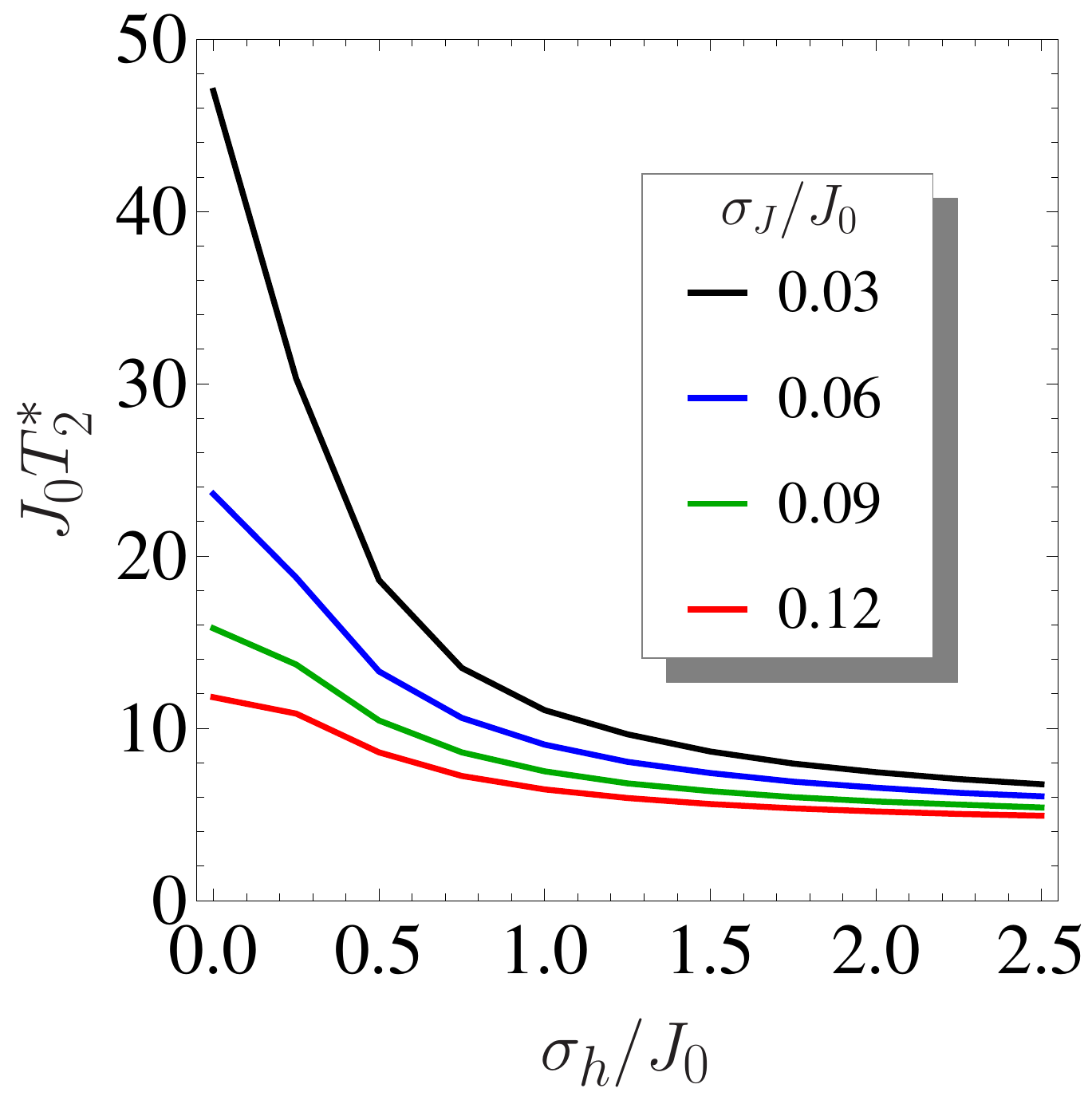}
\includegraphics[width=0.49\columnwidth]{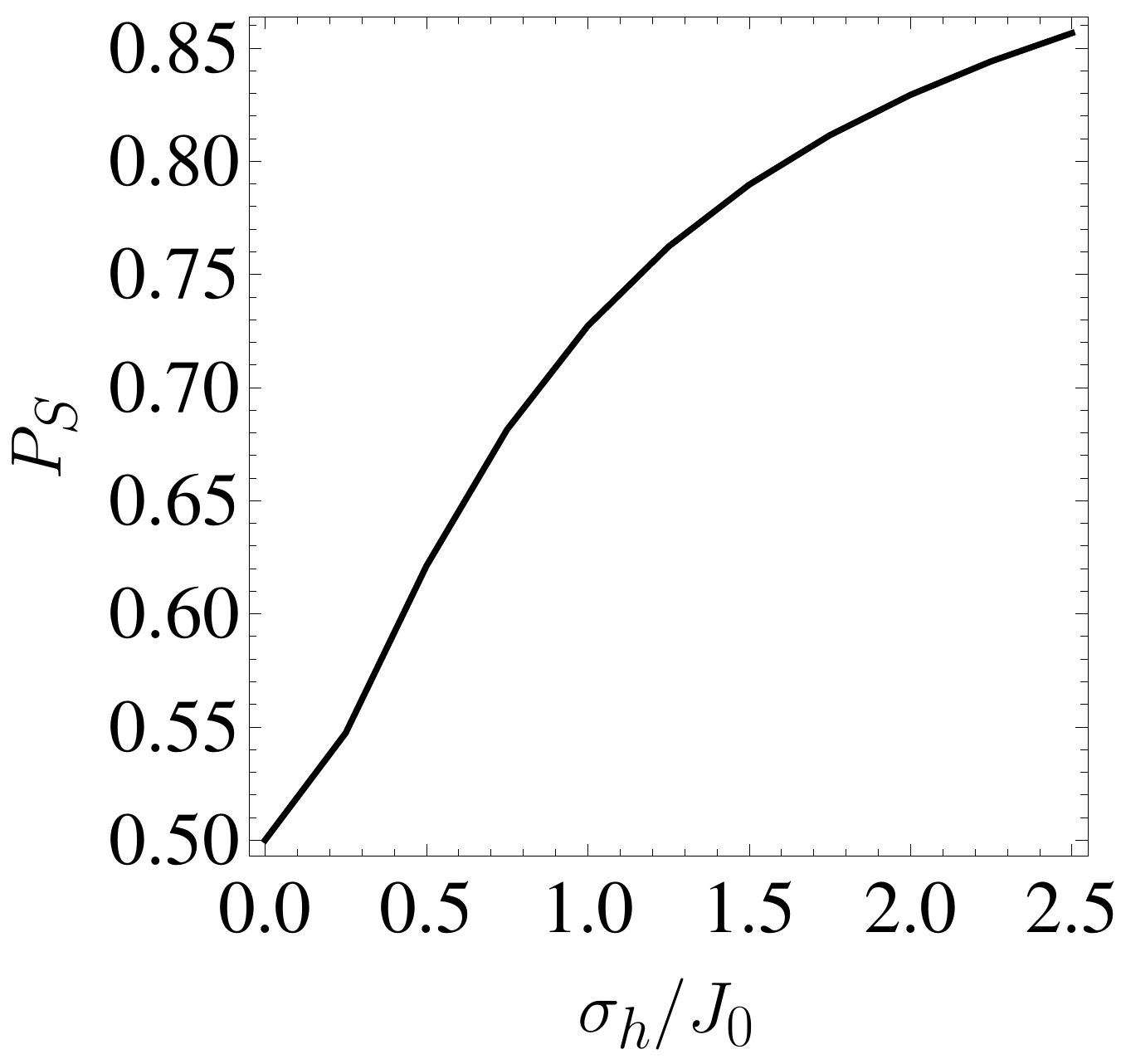}
\caption{(Top left) Plots of the (dimensionless) coherence time, $J_0 T_2^{\ast}$, as a function of $\sigma_J/J_0$
for several values of $\sigma_h/J_0$.  (Top right) Plots of the same as a function of $\sigma_h/J_0$ for several
values of $\sigma_J/J_0$.  (Bottom) Plot of the return probability as a function of $\sigma_h/J_0$ for $\sigma_J=0.03J_0$.
We only present this value because the results for larger values of $\sigma_J$ do not differ significantly from
this curve.}
\label{Fig:T2S_Slices}
\end{figure}

Finally, we give plots of our general results for $T_2^{\ast}$ and $P_S$ as functions of $\sigma_h/J_0$
and $\sigma_J/J_0$.  We first present such plots over the region, $0\leq\sigma_h/J_0\leq 2.5$ and
$0\leq\sigma_J/J_0\leq 0.15$, in Fig.\ \ref{Fig:Results_10_100} since these are within the expected
regime of experimental interest in GaAs and Si systems.  We note that typically the charge
noise is much weaker than the field noise except for isotopically pure Si where the two may be comparable
in magnitude with the field noise arising simply from fluctuations in the applied magnetic field in
contrast to GaAs where the main source of field noise is Overhauser nuclear field fluctuations.  We find that the return probability
as a function of $\sigma_h/J_0$ does not change noticeably if we fix $\sigma_J/J_0$ to a nonzero value,
rather than to zero.  We also see the clear trends in $T_2^{\ast}$ that our results for the $\sigma_h=0$ and
$\sigma_J=0$ limits imply, namely, that it decreases if we increase either type of disorder, but increasing
$\sigma_J/J_0$ has a quantitatively stronger effect than $\sigma_h/J_0$.  We indicate on these plots the strength of the
disorder present in the experiments described in Ref.\ \onlinecite{MartinsPRL2016}.  We also present
``slices'' of the plot of the coherence time for constant $\sigma_h/J_0$ and $\sigma_J/J_0$ in Fig.\ \ref{Fig:T2S_Slices}.

\subsection{``Quality factor''}
\begin{figure}[ht]
\includegraphics[width=0.5\columnwidth]{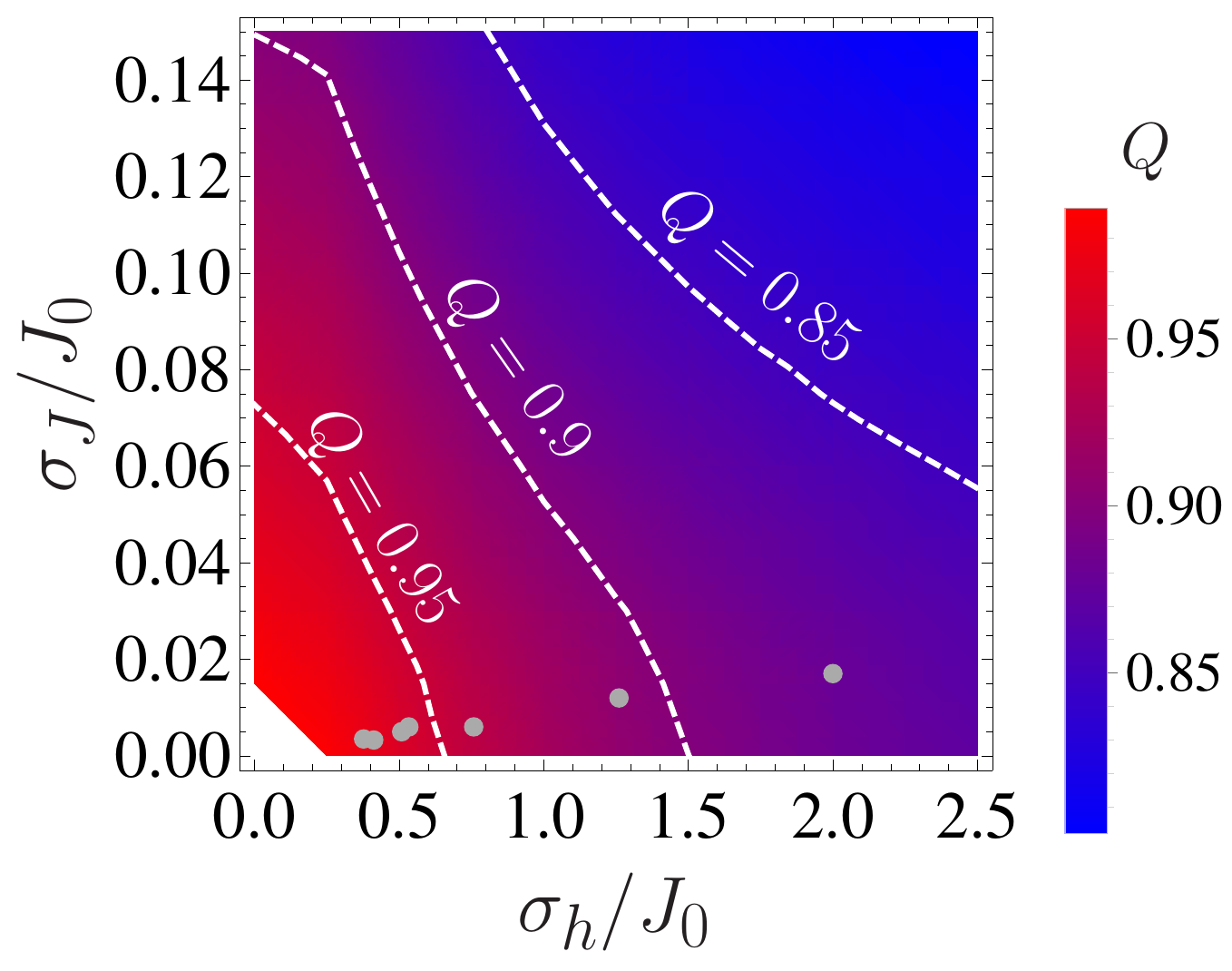}
\caption{Plot of the quality factor $Q$ as a function of $\sigma_h/J_0$ and $\sigma_J/J_0$ over the domain,
$0\leq\sigma_h/J_0\leq 2.5$ and $0\leq\sigma_J/J_0\leq 0.15$. The dashed lines indicates points at which $Q=0.85$,
$Q=0.9$, and $Q=0.95$.  The gray dots represent the strength of the noise present in the experiments described
in Ref.\ \onlinecite{MartinsPRL2016}.}
\label{Fig:F_10_100}
\end{figure}

\begin{figure}[ht]
\includegraphics[width=0.49\columnwidth]{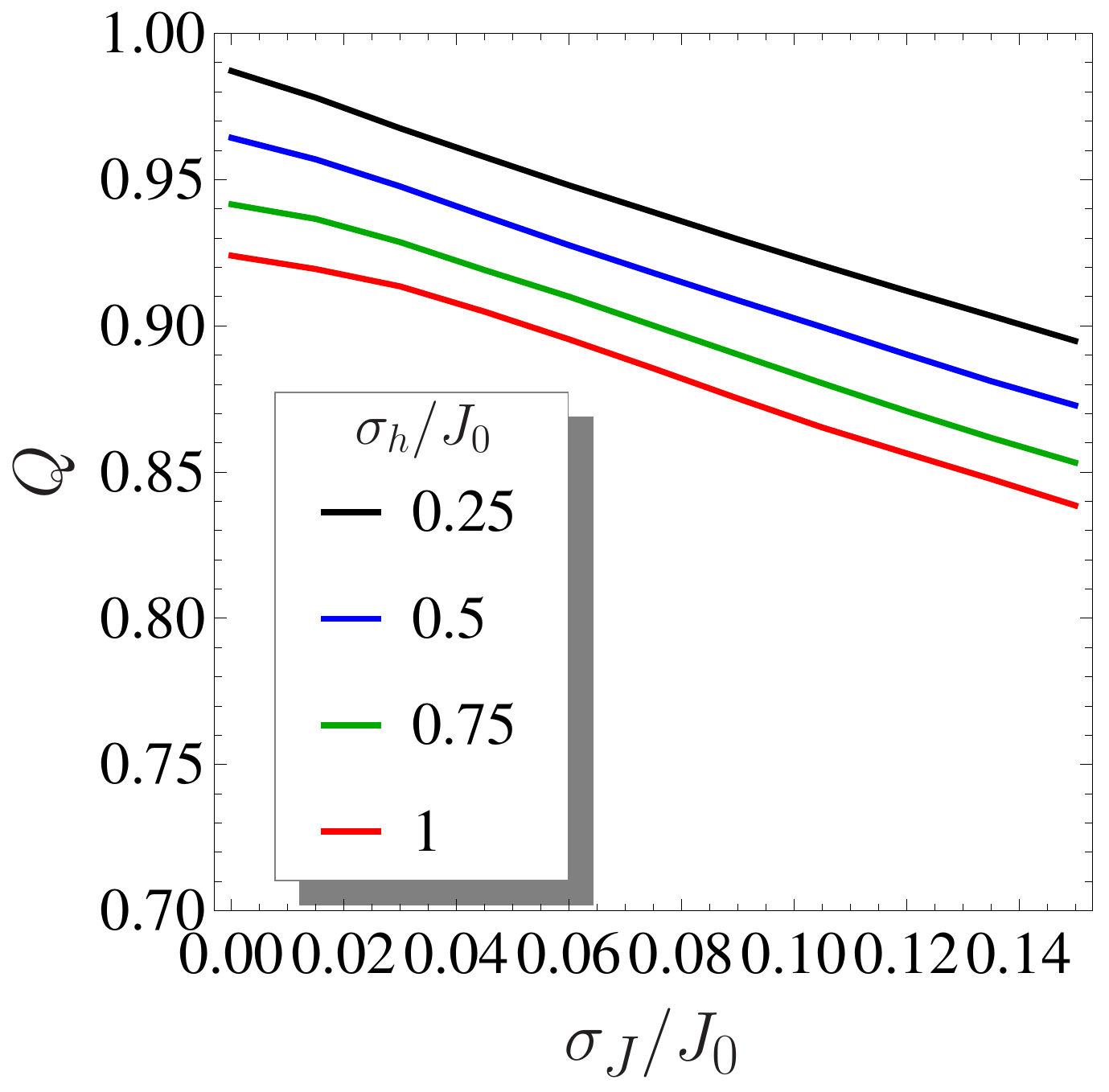}
\includegraphics[width=0.49\columnwidth]{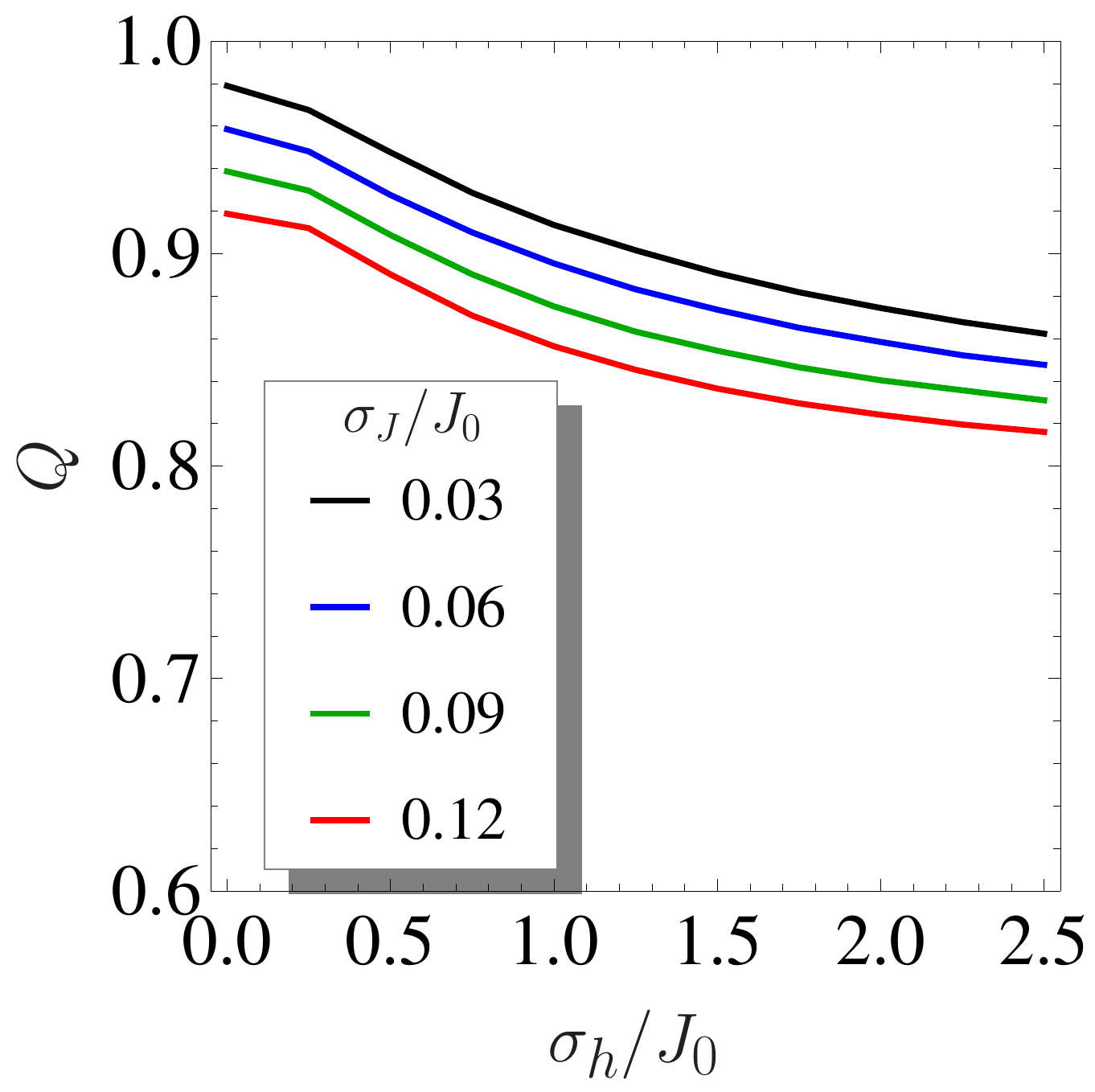}
\caption{(Left) Plot of the quality factor $Q$ as a function of $\sigma_J/J_0$ for several values of $\sigma_h/J_0$.
(Right) Plot of the same as a function of $\sigma_h/J_0$ for several values of $\sigma_J/J_0$.}
\label{Fig:F_Slices}
\end{figure}

We now present our results for the coherence time in an alternate form, in terms of a ``quality factor'', which we may
directly obtain from our results for $T_2^{\ast}$, and which will prove useful in our discussions about entanglement fidelity
below.  We define this quality factor operationally as $Q=e^{-1/J_0 T_2^{\ast}}$.  We first plot it over the domain of current
physical interest,  $0\leq\sigma_h/J_0\leq 2.5$ and $0\leq\sigma_J/J_0\leq 0.15$, in Fig.\ \ref{Fig:F_10_100}; we also show
``slices'' of this plot for constant $\sigma_h/J_0$ and $\sigma_J/J_0$ in Fig.\ \ref{Fig:F_Slices}.  As with our $T_2^{\ast}$ results,
we also show where the experimental data of Ref.\ \onlinecite{MartinsPRL2016} fall within this region.  We see that some of the
experimental data already have quality factors exceeding $0.95$.

\begin{figure}[ht]
\includegraphics[width=0.49\columnwidth]{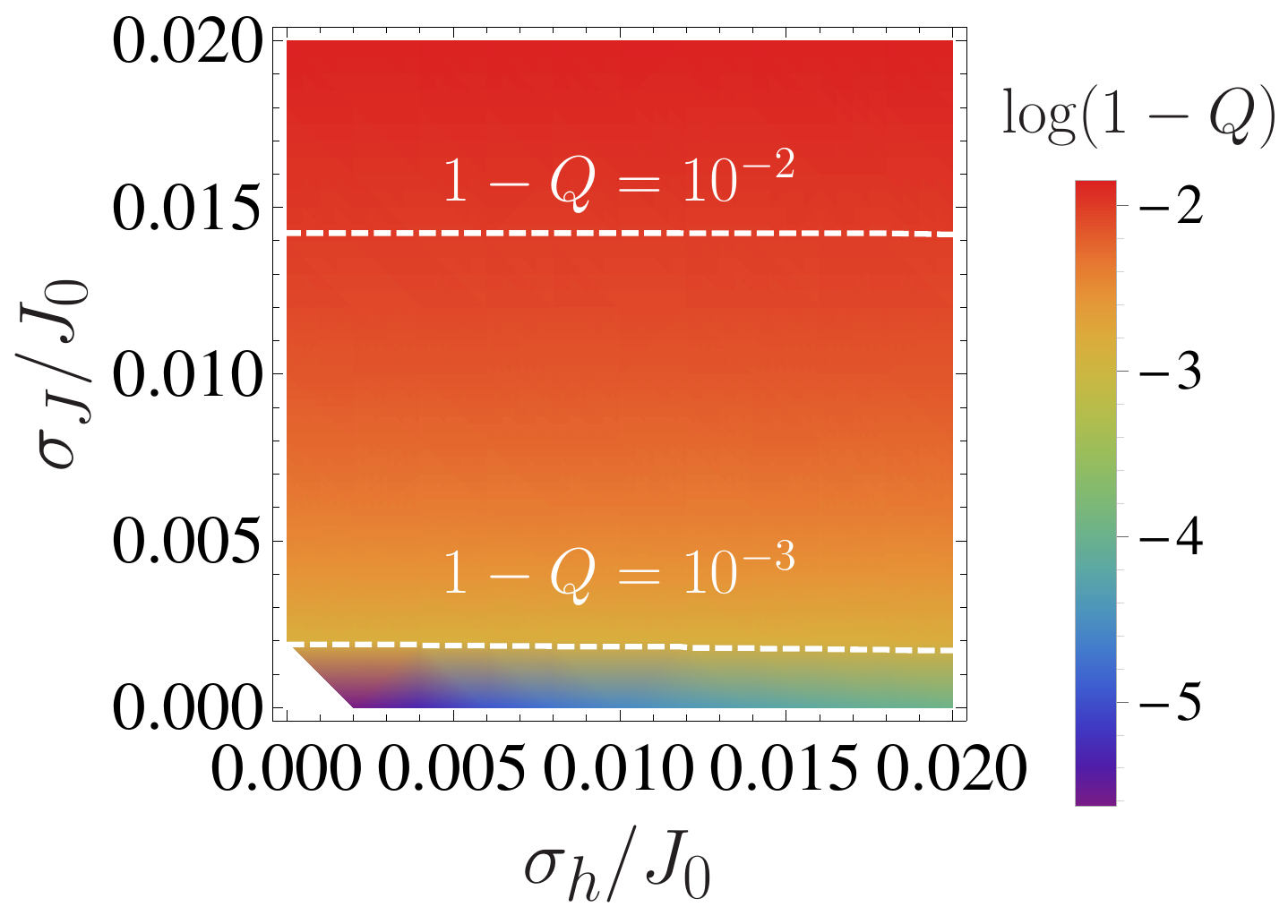}
\includegraphics[width=0.49\columnwidth]{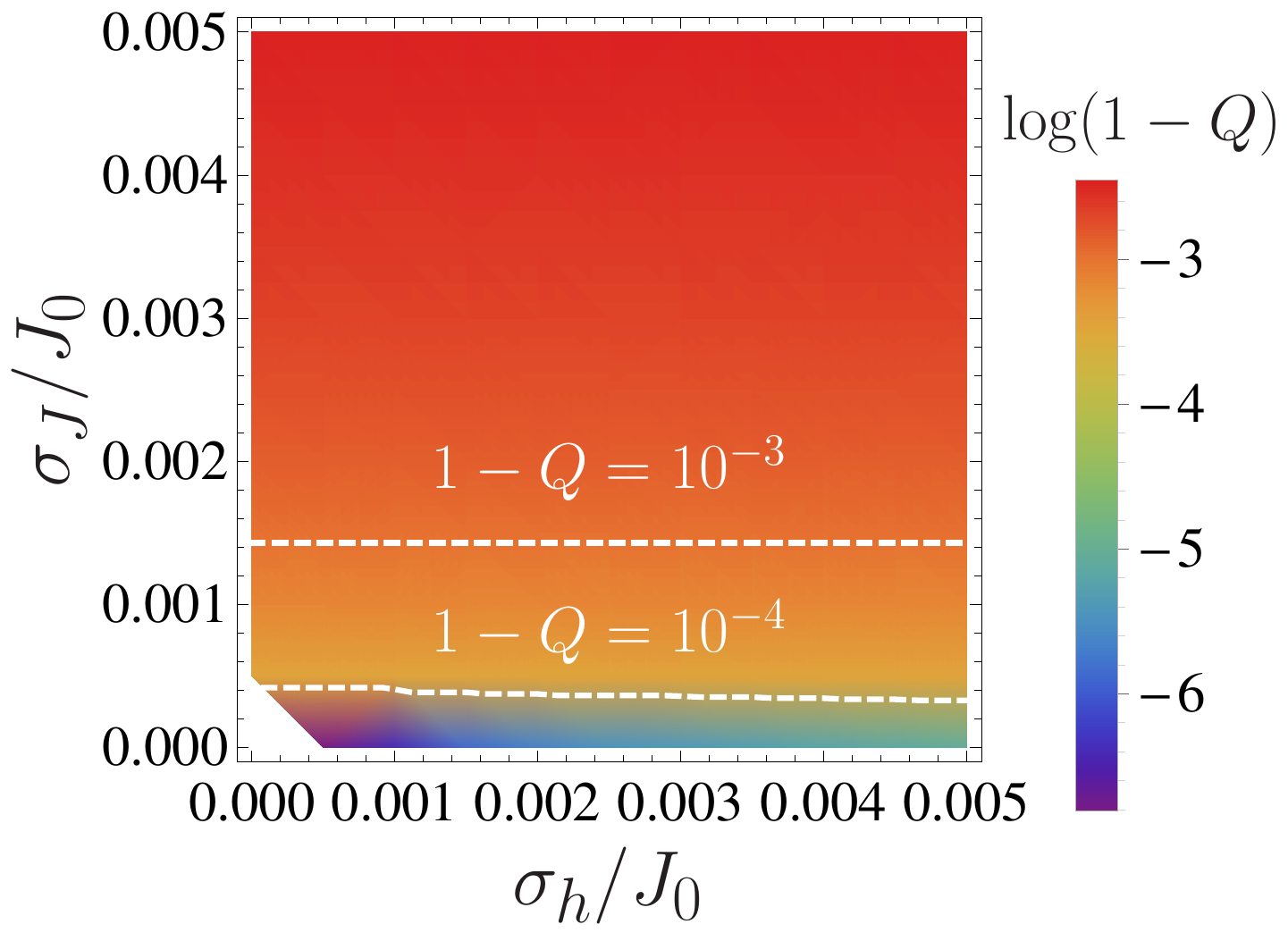}
\includegraphics[width=0.49\columnwidth]{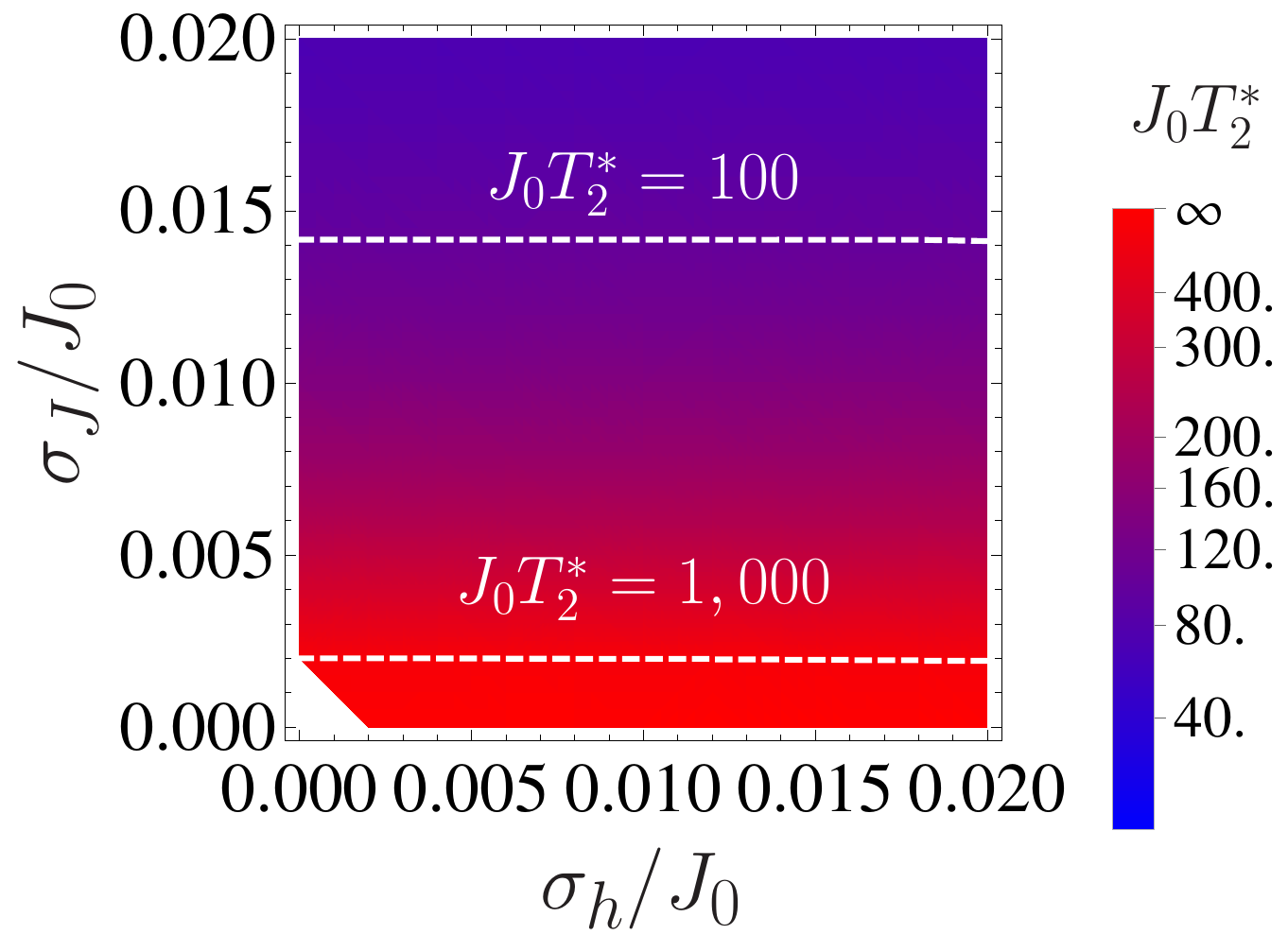}
\includegraphics[width=0.49\columnwidth]{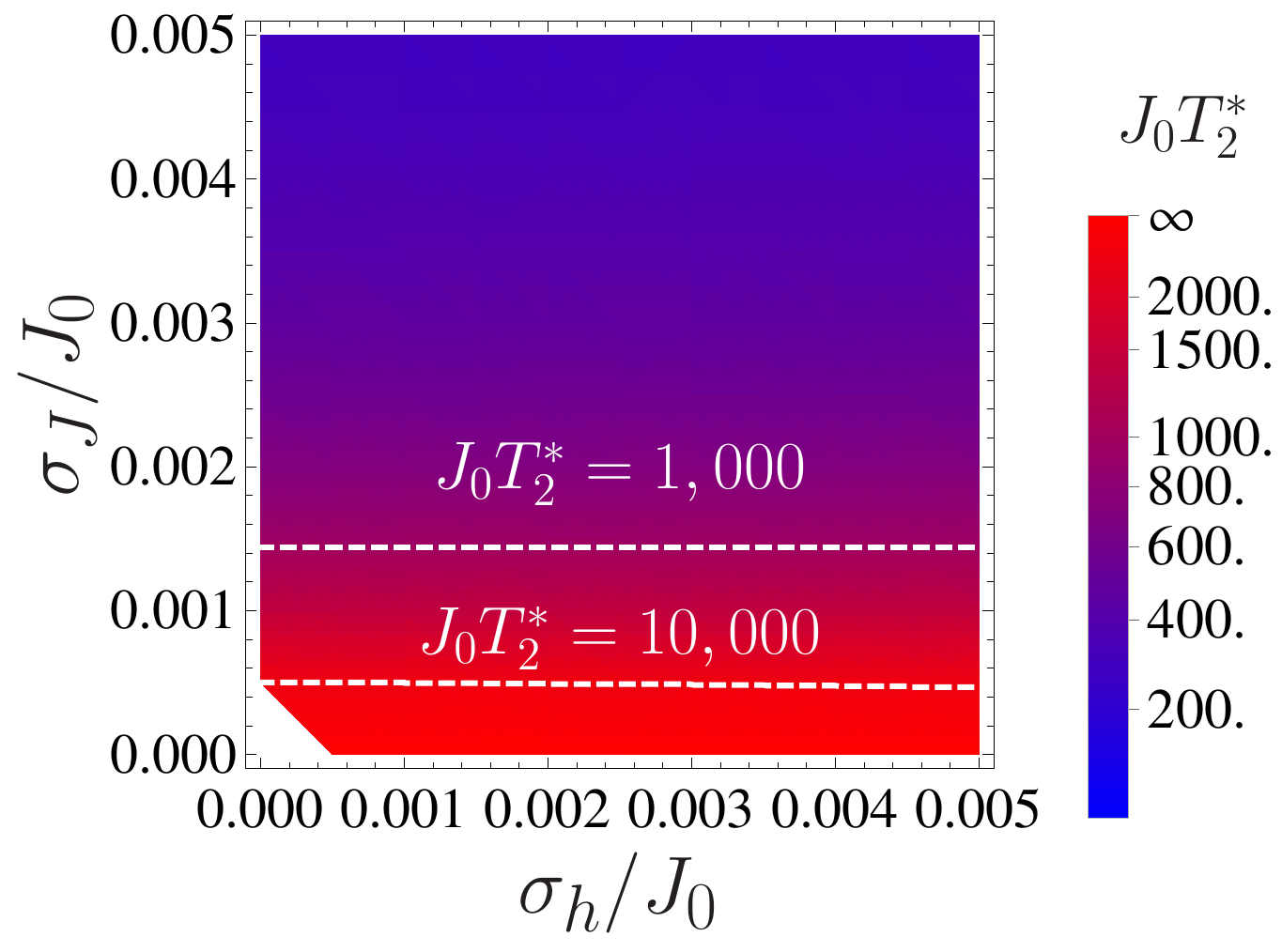}
\caption{Plot of the (common) logarithm of $\log(1-Q)$, where $Q$ is the quality factor, over the regions, $0\leq\sigma_h/J_0\leq0.02$
and $0\leq\sigma_J/J_0\leq0.02$ (top left) and $0\leq\sigma_h/J_0\leq0.005$ and $0\leq\sigma_J/J_0\leq0.005$ (top right).
We also include plots of the (dimensionless) coherence time, $J_0 T_2^{\ast}$, over the same regions on the bottom left and
bottom right.}
\label{Fig:F_Zoomed}
\end{figure}

Finally, we provide plots over smaller regions in Fig.\ \ref{Fig:F_Zoomed}, namely,  $0\leq\sigma_h/J_0\leq0.02$ and
$0\leq\sigma_J/J_0\leq0.02$, and $0\leq\sigma_h/J_0\leq0.005$ and $0\leq\sigma_J/J_0\leq0.005$.  In this case, due
to the fact that the quality factors are close to $1$ (at least $0.986$), we instead plot $\log(1-Q)$ for visual clarity,
where $\log$ is the common logarithm.  The results shown in Fig.\ \ref{Fig:F_Zoomed} provide the constraints on the noise
that must be achieved in future experiments in order for the semiconductor spin quantum computing platform to approach
the quantum error correction threshold. 

\section{Fidelity}
We now determine the fidelity for producing the two maximally entangled states, $\ket{ME_1}$ and $\ket{ME_2}$,
from the unentangled state, $\ket{\uparrow\downarrow}$.  The fidelity $F_S$ of preparation of a state
$\ket{S}$ from some initial state $\ket{I}$ is defined as the disorder-averaged probability that, upon measuring
the state of the system after performing the operation $O$ that would ideally transform $\ket{I}$ into $\ket{S}$,
we will actually find it in the state $\ket{S}$\cite{NielsenBook,MagesanPRA2012}:
\begin{equation}
F_S=\left [\left |\braket{S|O|I}\right |^2\right ]_{\alpha}.
\end{equation}
We should emphasize that this definition of the fidelity is dependent on the initial state of the system and thus
should not be thought of as a two-qubit entangling gate fidelity. It is more properly thought of as a state fidelity that characterizes 
the role of noise and decoherence in the generation of entanglement starting from a specific, unentangled initial state.
However, we expect the two fidelities (our state fidelity and the entangling gate fidelity) not to be very different
from each other; in particular, they should manifest similar qualitative dependence on the noise.  Our goal is to 
illustrate the role of both changes in the steady-state return probability and in the intrinsic decoherence time in
reducing these entanglement fidelities, assuming that we start from a maximally unentangled state.  We will now apply
this definition to $\ket{ME_1}$ and $-\ket{ME_2}$, which are the states that we would ideally obtain under simple time
evolution by $t=\pi/2J_0$ and $t=3\pi/2J_0$, respectively.  We begin with the state, $\ket{ME_1}$.  We find that the
probability at time $t$ for obtaining this state, starting from the initial state, $\ket{\uparrow\downarrow}$, is
\begin{equation}
P_{ME_1}(t)=\frac{1}{2}+\frac{J}{2\sqrt{J^2+(\delta h)^2}}\sin\left [\sqrt{J^2+(\delta h)^2}t\right ].
\end{equation}
The fidelity for the preparation of $\ket{ME_1}$ is then simply the disorder average of this probability at time $t=\pi/2J_0$:
\begin{equation}
F_{ME_1}=\left [P_{ME_1}\left (\frac{\pi}{2J_0}\right )\right ]_{\alpha}.
\end{equation}
Similarly, the probability for entering the state, $\ket{ME_2}$, at time $t$ is
\begin{equation}
P_{ME_2}(t)=\frac{1}{2}-\frac{J}{2\sqrt{J^2+(\delta h)^2}}\sin\left [\sqrt{J^2+(\delta h)^2}t\right ],
\end{equation}
and the fidelity of preparation of this state is simply the disorder average of this probability at time
$t=3\pi/2J_0$:
\begin{equation}
F_{ME_2}=\left [P_{ME_2}\left (\frac{3\pi}{2J_0}\right )\right ]_{\alpha}.
\end{equation}
We now evaluate these averages, first in the $\sigma_h=0$ and $\sigma_J=0$ limits, and then for general
disorder strengths.

\subsection{$\sigma_h=0$ limit}
\begin{figure}[ht]
\includegraphics[width=0.49\columnwidth]{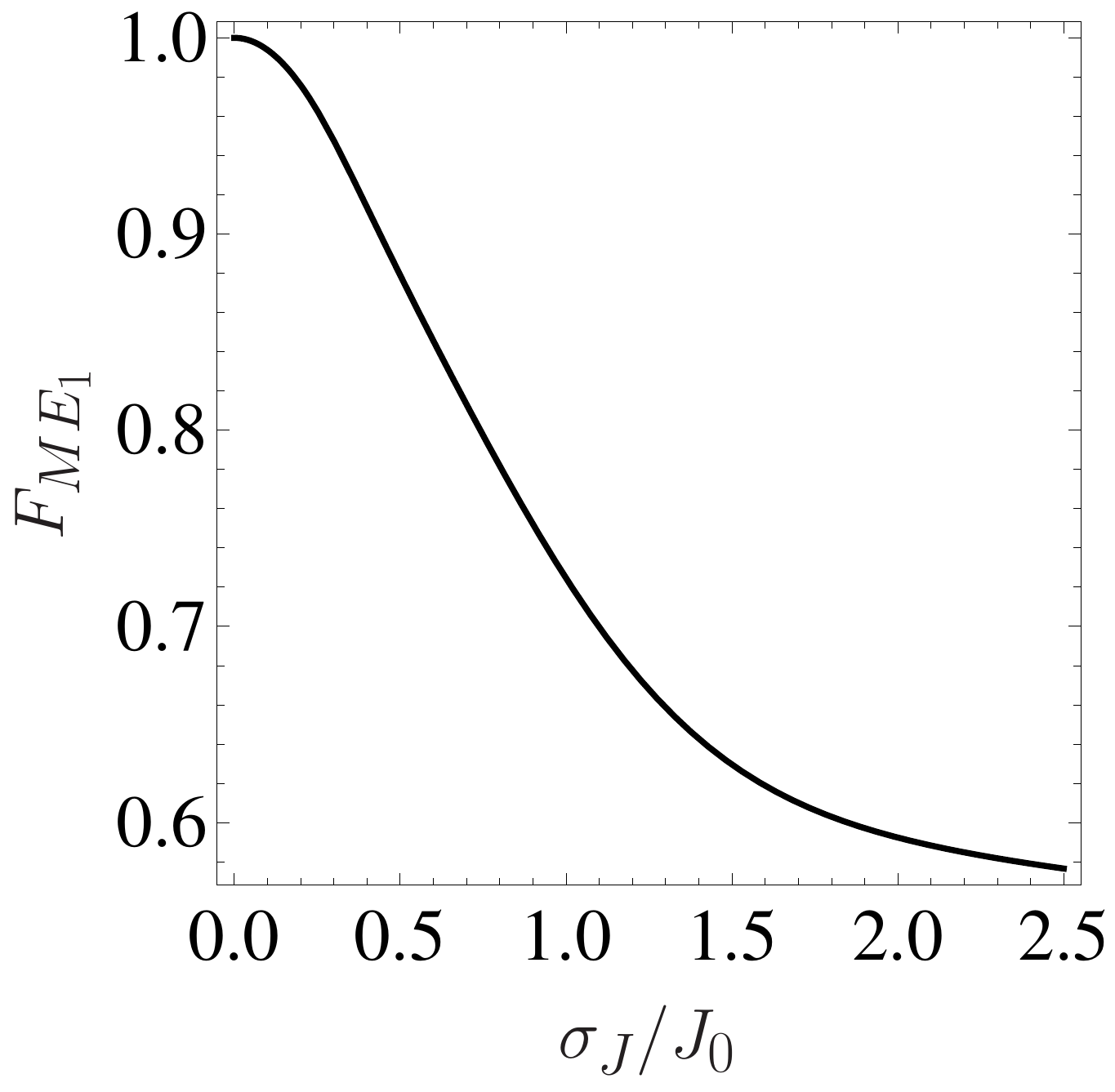}
\includegraphics[width=0.49\columnwidth]{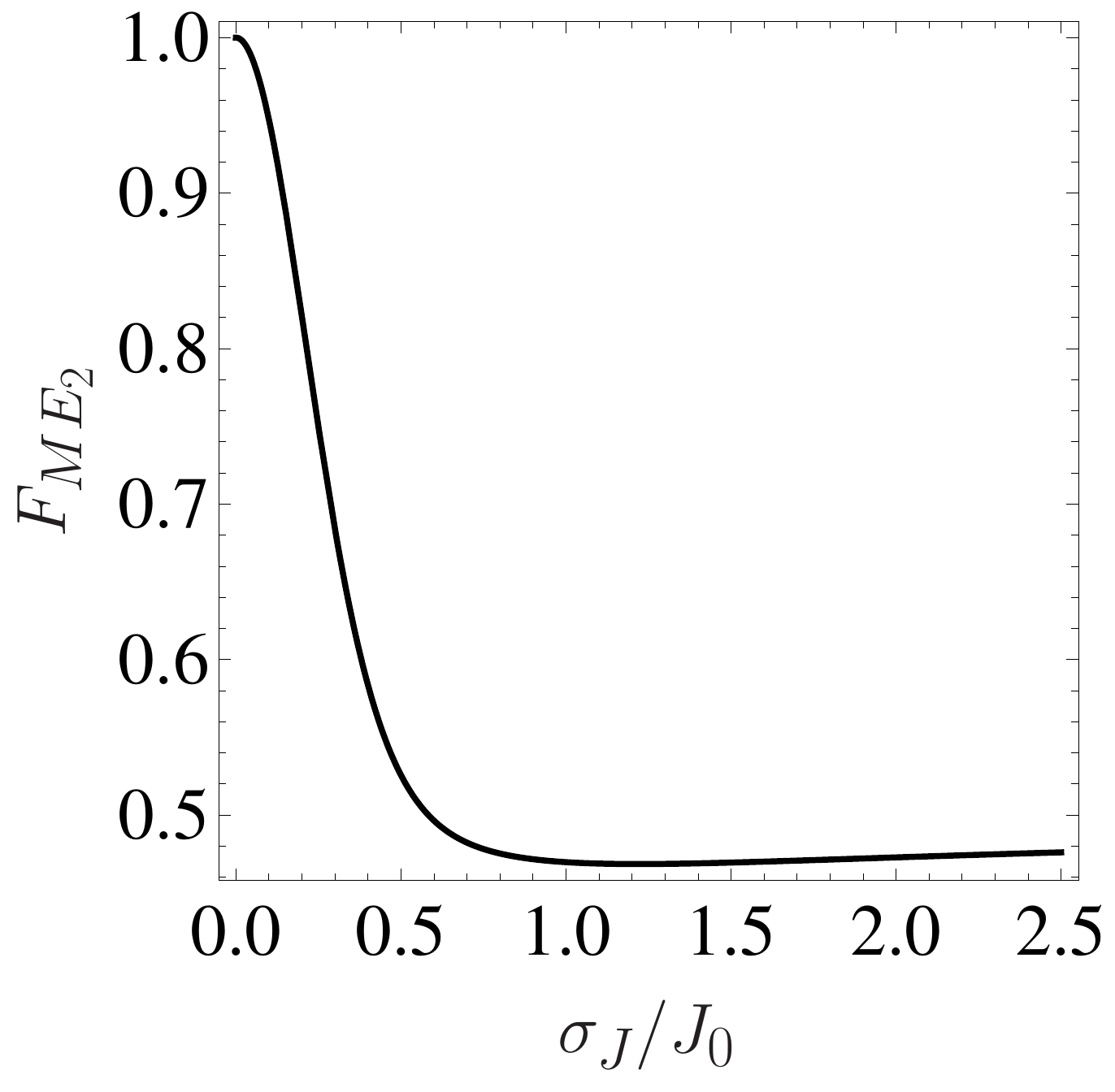}
\caption{Plots of $F_{ME_1}$ (left) and $F_{ME_2}$ (right) as a function of $\sigma_J/J_0$ for $\sigma_h=0$.}\label{Fig:Fidelity_noSigh}
\end{figure}

We first consider the $\sigma_h=0$ limit.  Here, we find that we can derive closed-form analytical
expressions for the fidelities.  We may write $F_{ME_1}$ as
\begin{eqnarray}
F_{ME_1}&=&\frac{1}{2}+\frac{1}{\sigma_J\sqrt{2\pi}}\frac{1}{1+\mbox{erf}(J_0/\sigma_J\sqrt{2})} \cr
&\times&\mbox{Im}\left [\int_0^{\infty}dJ\,e^{-(J-J_0)^2/2\sigma_J^2+i\pi J/2J_0}\right ].
\end{eqnarray}
This integral may be evaluated in terms of the error function; we obtain
\begin{eqnarray}
F_{ME_1}&=&\frac{1}{2}+\frac{1}{2[1+\mbox{erf}(J_0/\sigma_J\sqrt{2})} \cr
&\times& e^{-\pi^2\sigma_J^2/8J_0^2}\mbox{Re}\left [1+\mbox{erf}\left (\frac{J_0}{\sigma_J\sqrt{2}}+i\frac{\pi\sigma_J}{2\sqrt{2}J_0}\right )\right ]. \nonumber \\
\end{eqnarray}
A similar calculation for $F_{ME_2}$ yields
\begin{eqnarray}
F_{ME_2}&=&\frac{1}{2}+\frac{1}{2[1+\mbox{erf}(J_0/\sigma_J\sqrt{2})} \cr
&\times& e^{-9\pi^2\sigma_J^2/8J_0^2}\mbox{Re}\left [1+\mbox{erf}\left (\frac{J_0}{\sigma_J\sqrt{2}}+i\frac{3\pi\sigma_J}{2\sqrt{2}J_0}\right )\right ]. \nonumber \\
\end{eqnarray}
We provide plots of these fidelities in Fig.\ \ref{Fig:Fidelity_noSigh}.  We see that, while charge noise
reduces the fidelity of both operations, it has a greater effect on the fidelity for producing $\ket{ME_2}$.
This is not surprising, as this operation takes longer to execute than the one for $\ket{ME_1}$.
We also note that there is a value of $\sigma_J/J_0$ at which $F_{ME_2}$ actually goes below $\frac{1}{2}$,
then turns around and steadily increases, saturating at $\frac{1}{2}$, implying that the system is actually
slightly more likely to go into the state, $\ket{ME_1}$.  This is likely an artifact because we
truncate the distribution of exchange couplings to non-negative values only.

\subsection{$\sigma_J=0$ limit}
\begin{figure}[ht]
\includegraphics[width=0.49\columnwidth]{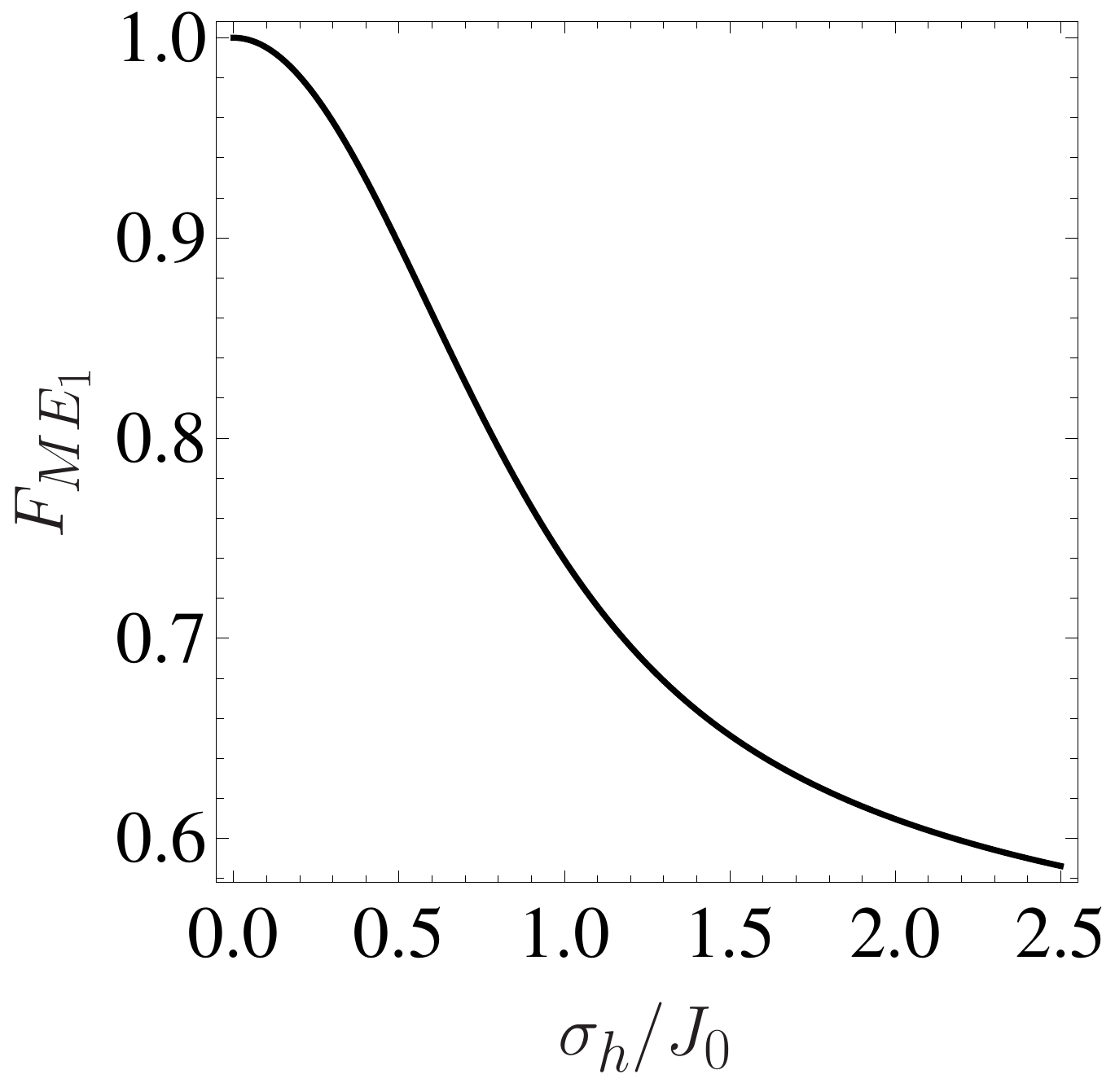}
\includegraphics[width=0.49\columnwidth]{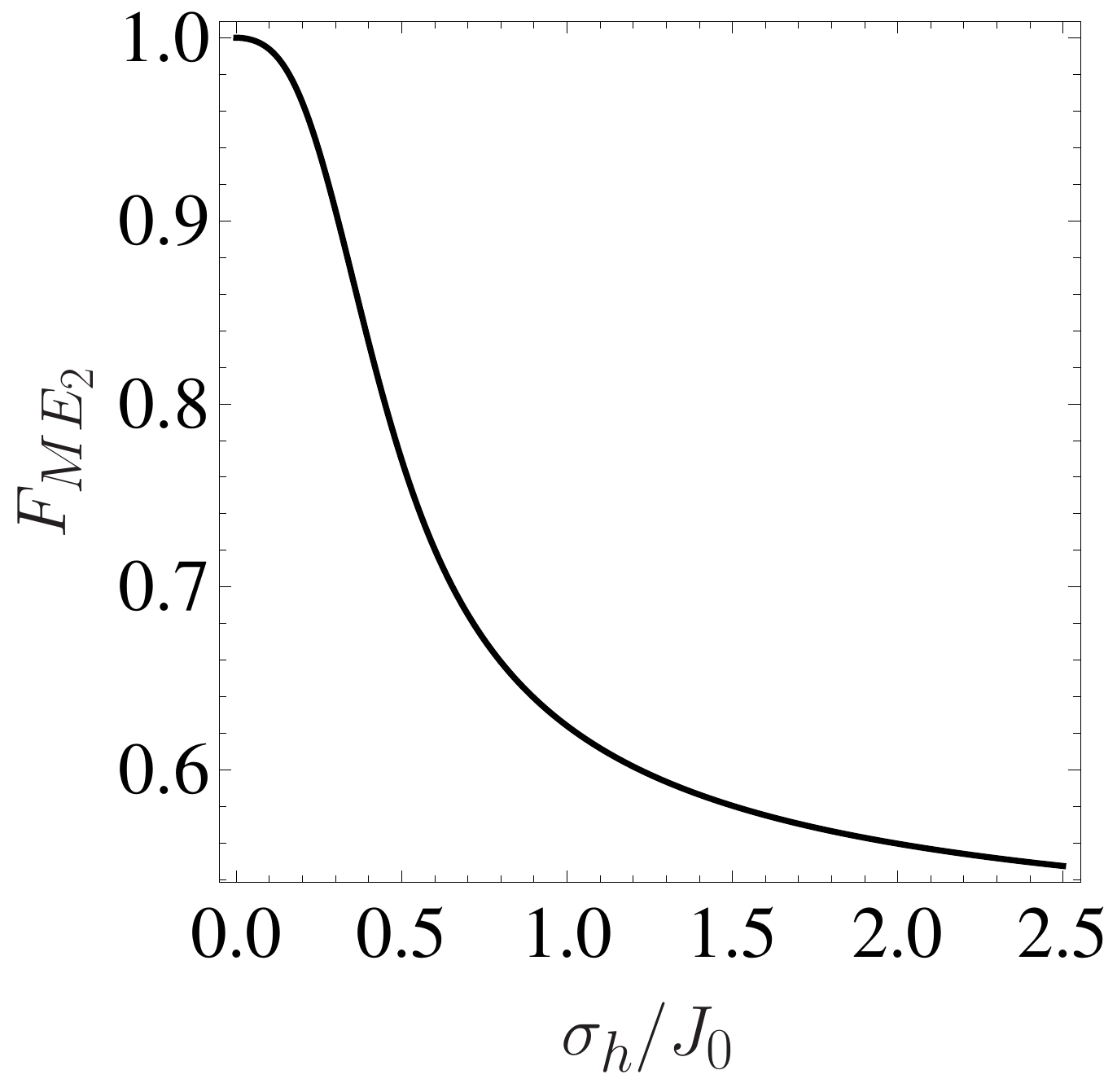}
\caption{Plots of $F_{ME_1}$ (left) and $F_{ME_2}$ (right) as a function of $\sigma_h/J_0$ for $\sigma_J=0$.}\label{Fig:Fidelity_noSigJ}
\end{figure}

We now consider the $\sigma_J=0$ limit.  Unlike the $\sigma_h=0$ limit, we cannot obtain analytical expressions
for the fidelities.  The expressions for the fidelities are
\begin{eqnarray}
F_{ME_1}&=&\frac{1}{2}+\frac{1}{2\sigma_h\sqrt{\pi}}\int_{-\infty}^{\infty}d(\delta h)\,\frac{J_0}{2\sqrt{J_0^2+(\delta h)^2}}e^{-(\delta h)^2/4\sigma_h^2} \cr
&\times&\sin\left [\frac{\pi}{2J_0}\sqrt{J_0^2+(\delta h)^2}\right ]
\end{eqnarray}
and
\begin{eqnarray}
F_{ME_2}&=&\frac{1}{2}-\frac{1}{2\sigma_h\sqrt{\pi}}\int_{-\infty}^{\infty}d(\delta h)\,\frac{J_0}{2\sqrt{J_0^2+(\delta h)^2}}e^{-(\delta h)^2/4\sigma_h^2} \cr
&\times&\sin\left [\frac{3\pi}{2J_0}\sqrt{J_0^2+(\delta h)^2}\right ].
\end{eqnarray}
We plot these expressions in Fig.\ \ref{Fig:Fidelity_noSigJ}.  We see a similar trend as before in the $\sigma_h=0$
limit, namely, that both fidelities are reduced, but $F_{ME_2}$ decreases more rapidly.  We note that field noise
seems to have as strong an effect on $F_{ME_1}$ as charge noise does, but that field noise has less of an effect on
$F_{ME_2}$ than charge noise.

\subsection{General results}
\begin{figure}[ht]
\includegraphics[width=0.49\columnwidth]{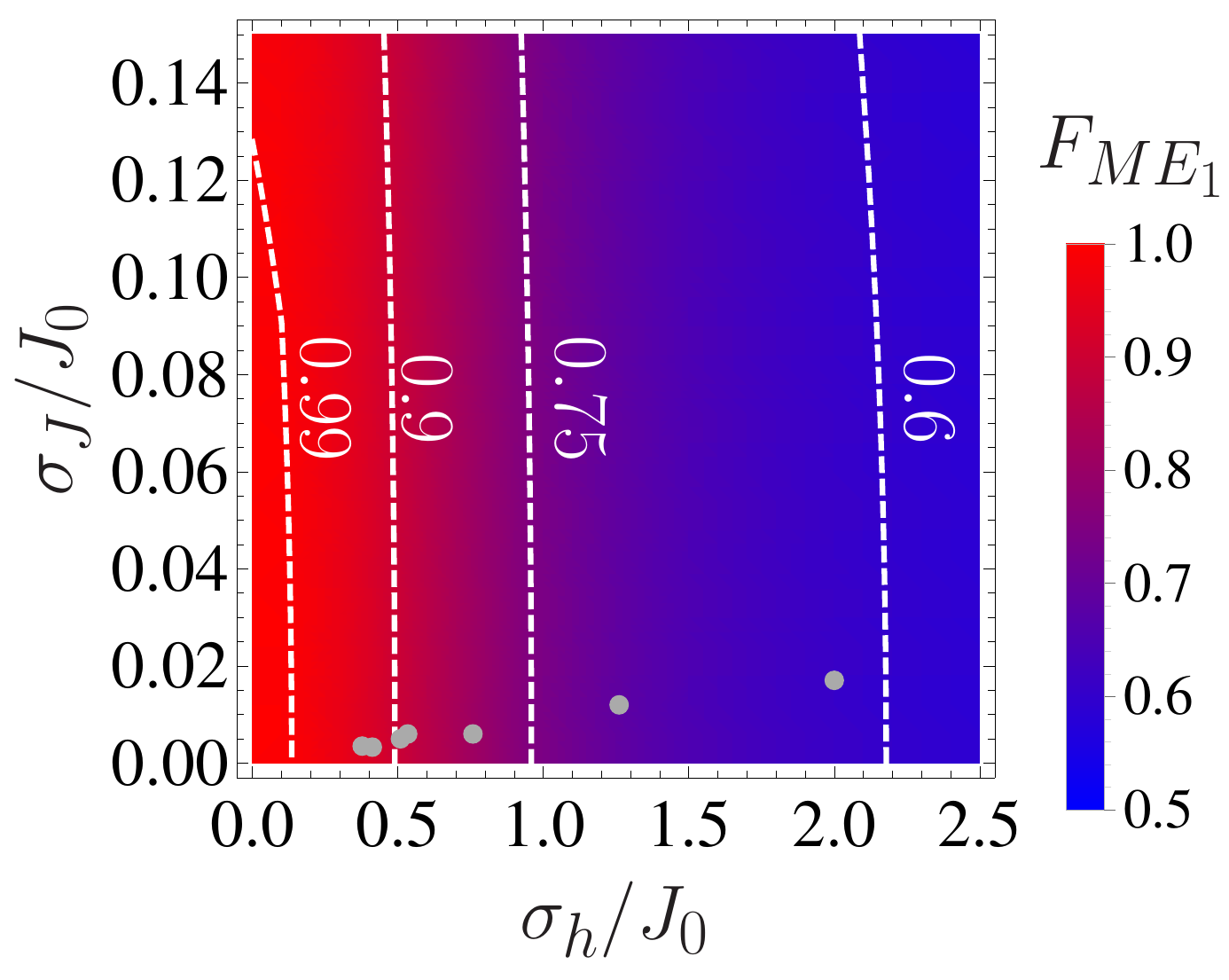}
\includegraphics[width=0.49\columnwidth]{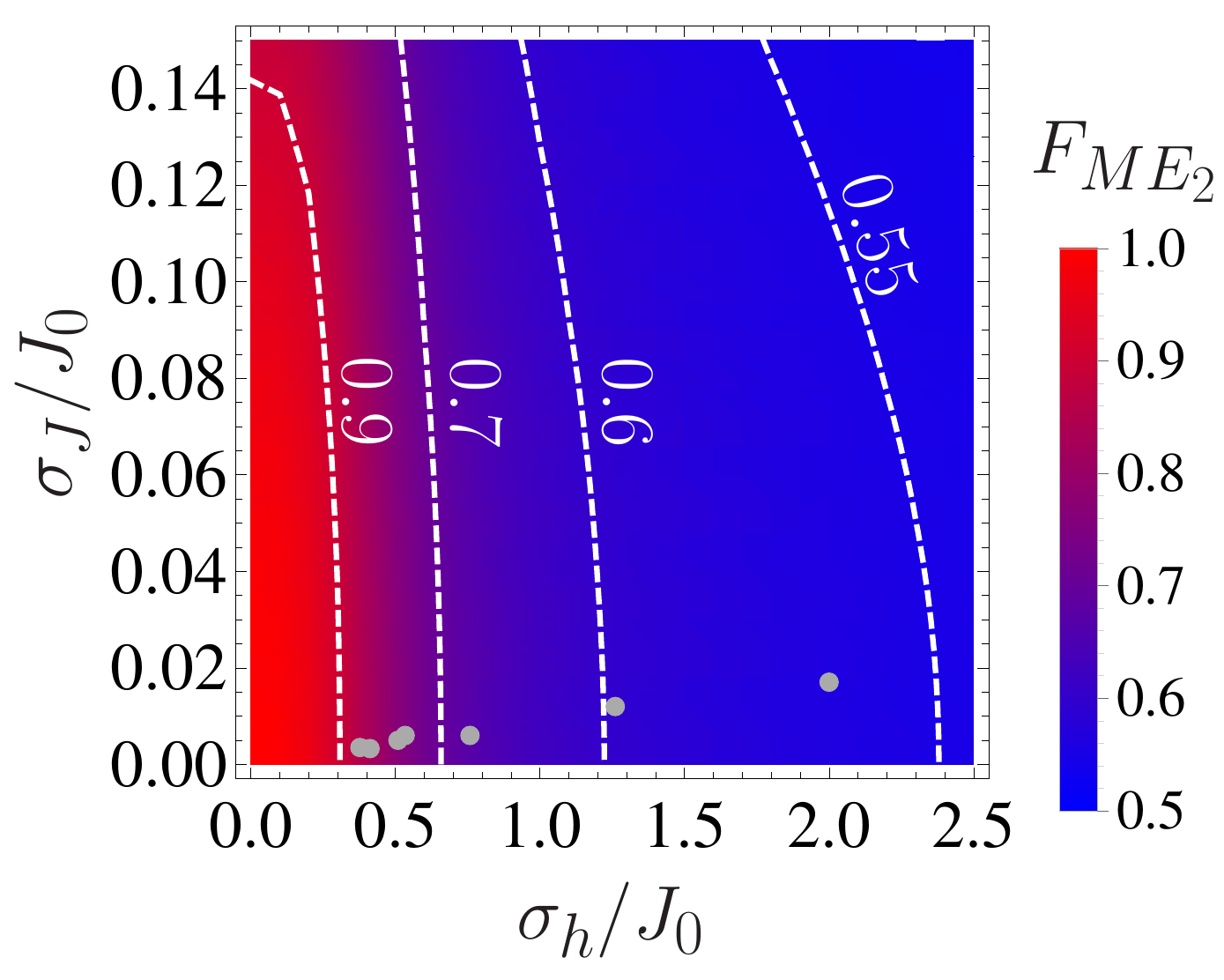}
\caption{Plots of $F_{ME_1}$ (left) and $F_{ME_2}$ (right) as a function of $\sigma_h$ and $\sigma_J$ over the region,
$0<\sigma_h/J_0\leq 2.5$ and $0<\sigma_J/J_0\leq 0.15$.  We also show contours (dashed lines) over which these fidelities
achieve specific values indicated on the plots.  The gray dots represent the strength of the noise present in the experiments
described in Ref.\ \onlinecite{MartinsPRL2016}.}\label{Fig:Fidelity_Gen}
\end{figure}

\begin{figure}[ht]
\includegraphics[width=0.49\columnwidth]{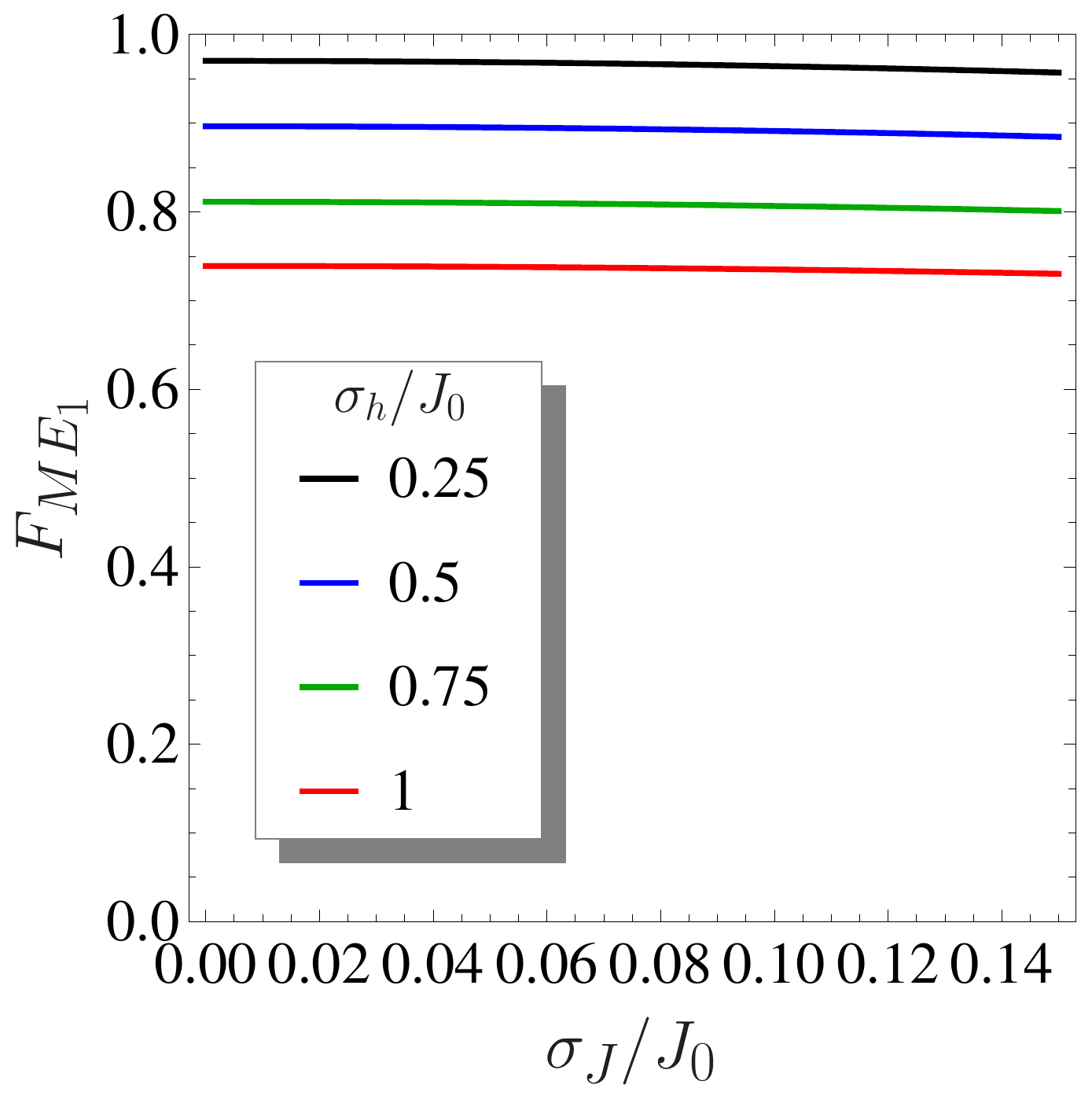}
\includegraphics[width=0.49\columnwidth]{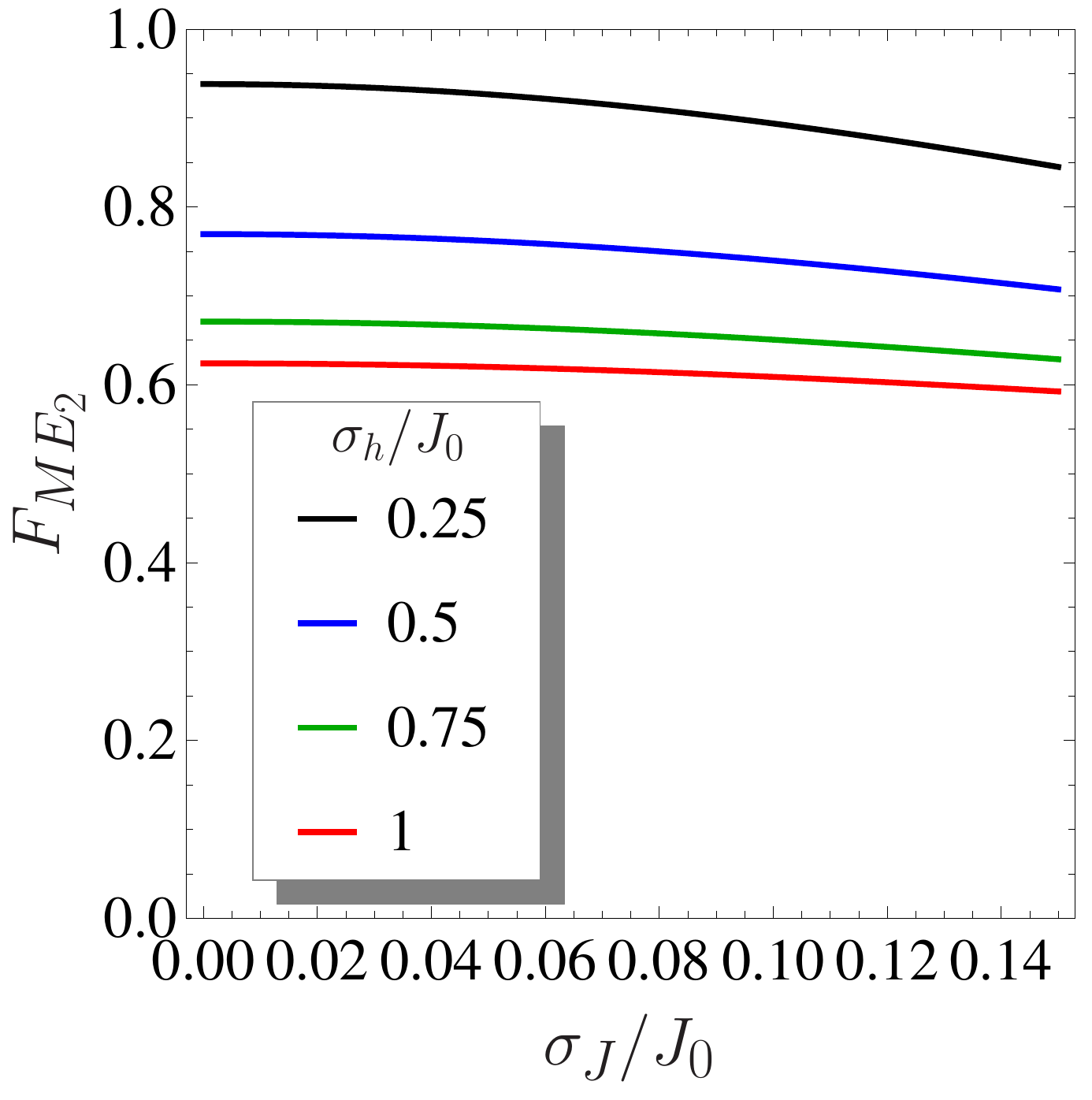}
\caption{Plots of the fidelities, $F_{ME_1}$ (left) and $F_{ME_2}$ (right) as a function of $\sigma_J/J_0$ for several values of
$\sigma_h/J_0$.}\label{Fig:Fidelity_ConstSigh}
\end{figure}

\begin{figure}[ht]
\includegraphics[width=0.49\columnwidth]{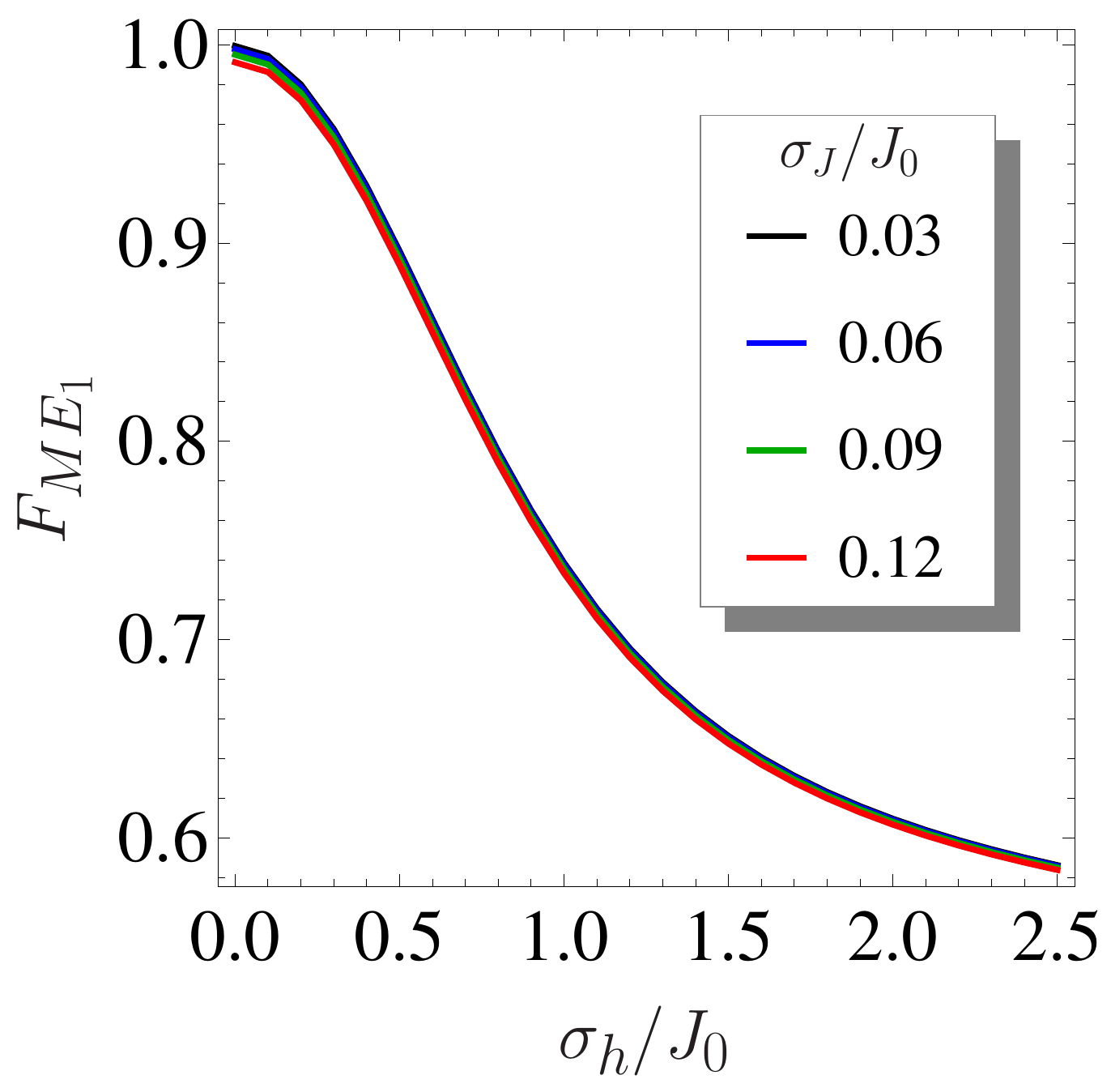}
\includegraphics[width=0.49\columnwidth]{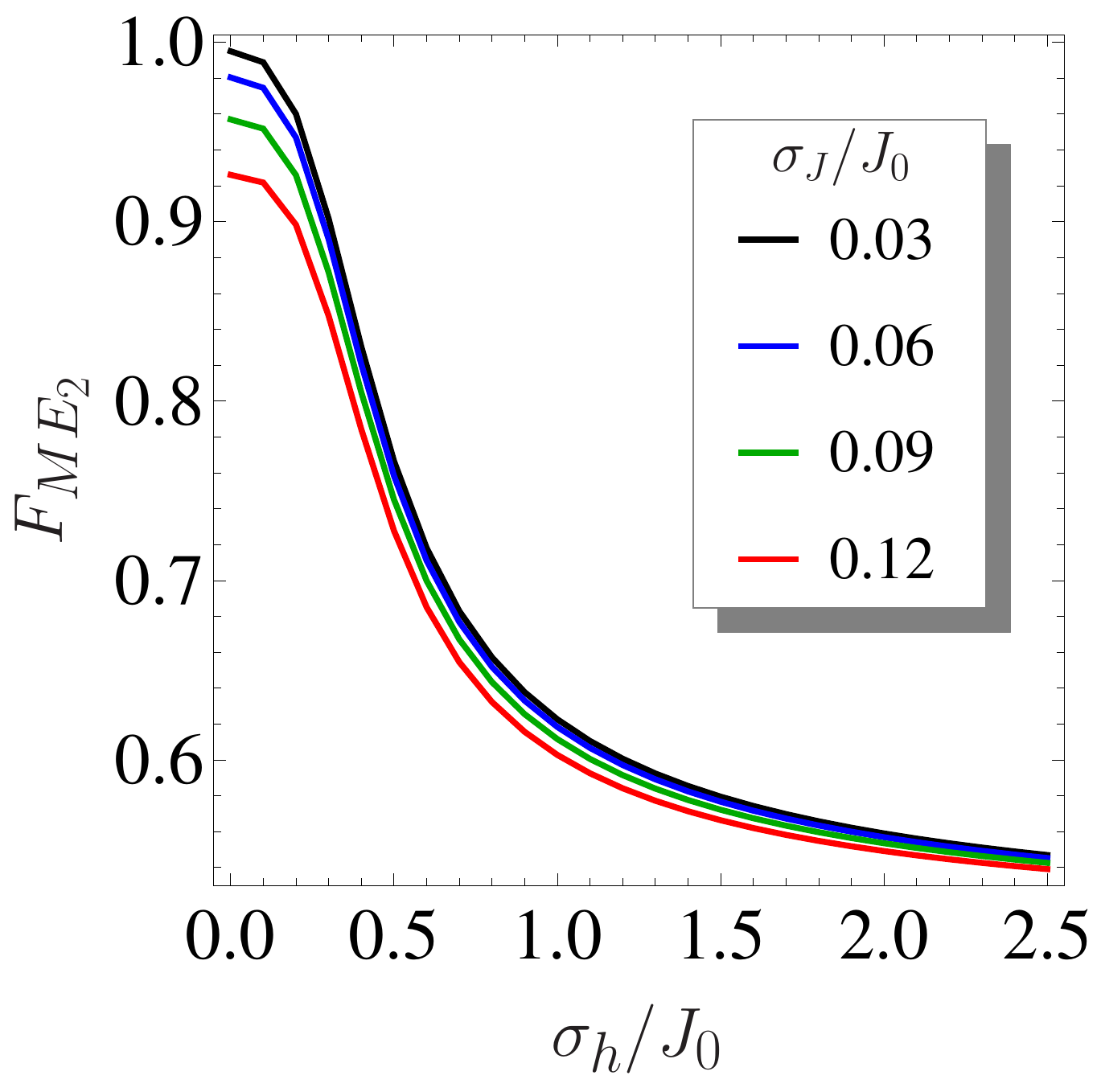}
\caption{Plots of the fidelities, $F_{ME_1}$ (left) and $F_{ME_2}$ (right) as a function of $\sigma_h/J_0$ for several values of
$\sigma_J/J_0$.}\label{Fig:Fidelity_ConstSigJ}
\end{figure}

\begin{figure}[ht]
\includegraphics[width=0.49\columnwidth]{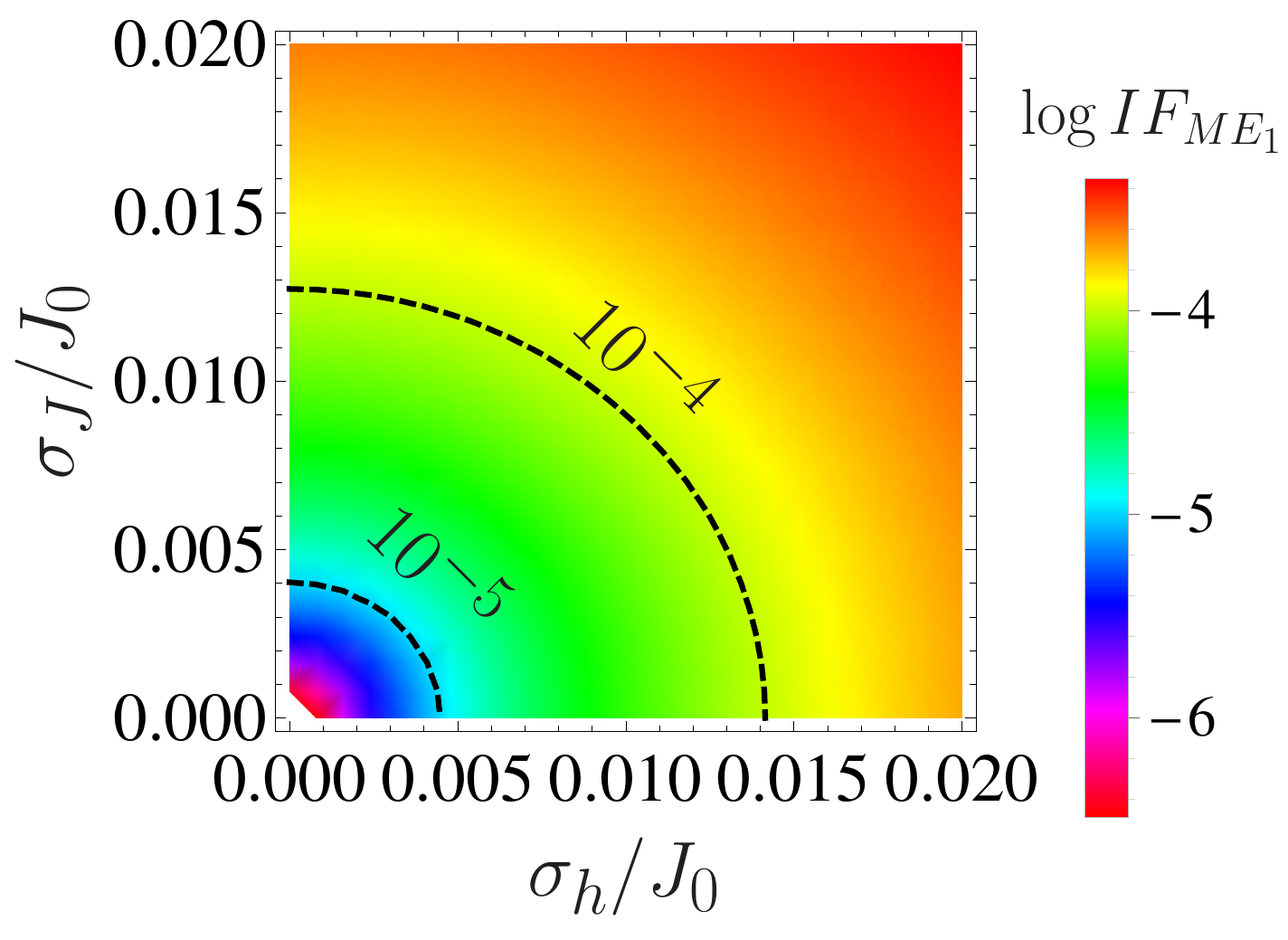}
\includegraphics[width=0.49\columnwidth]{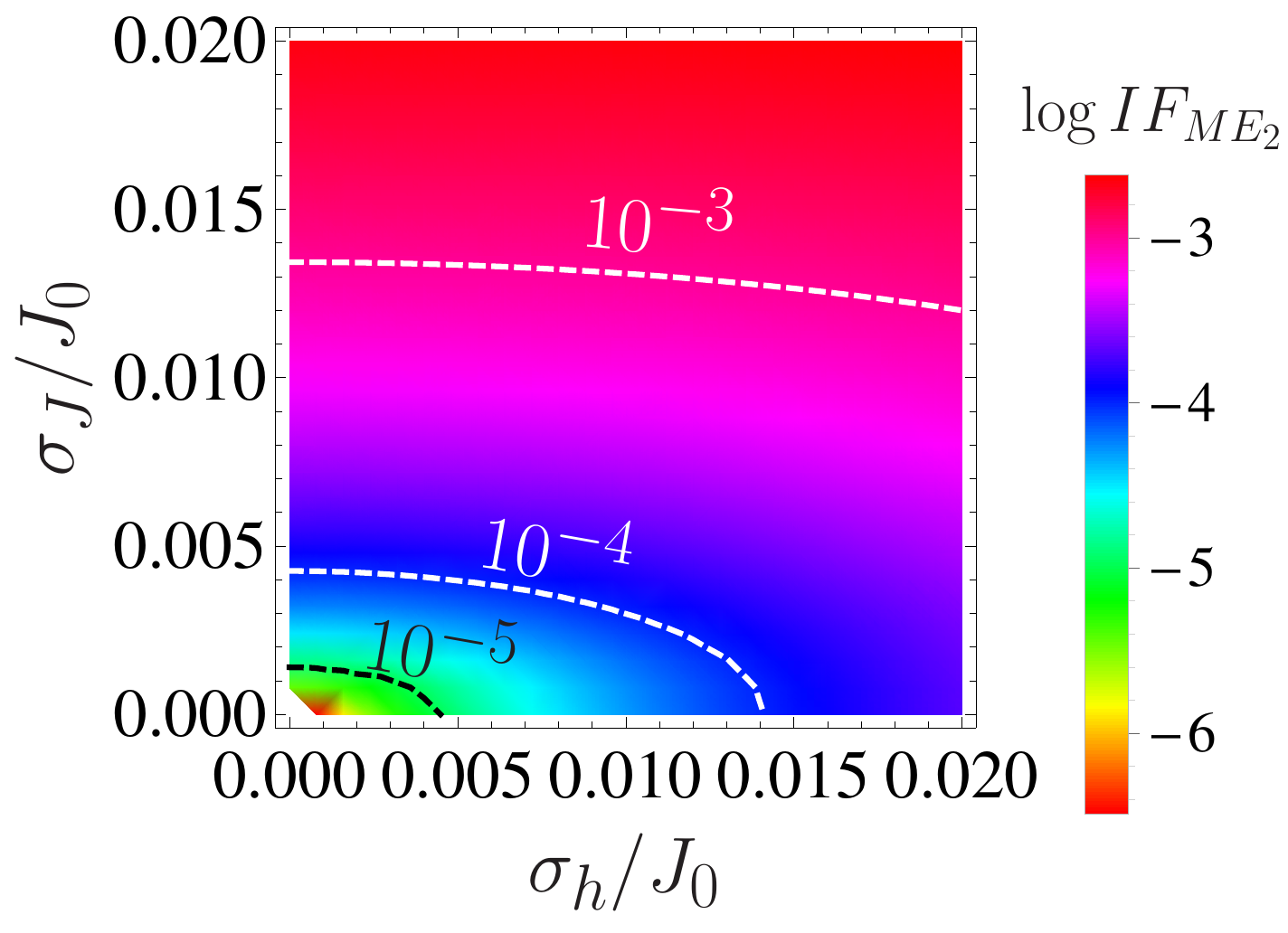}
\caption{Plots of the (common) logarithms of the infidelities, $IF_S=1-F_S$, over the region, $0<\sigma_h/J_0\leq 0.02$ and
$0<\sigma_J/J_0\leq 0.02$.  The labels on the contours are the values of the infidelities along said contours.}\label{Fig:Fidelity_Zoomed}
\end{figure}

\begin{figure}[ht]
\includegraphics[width=0.49\columnwidth]{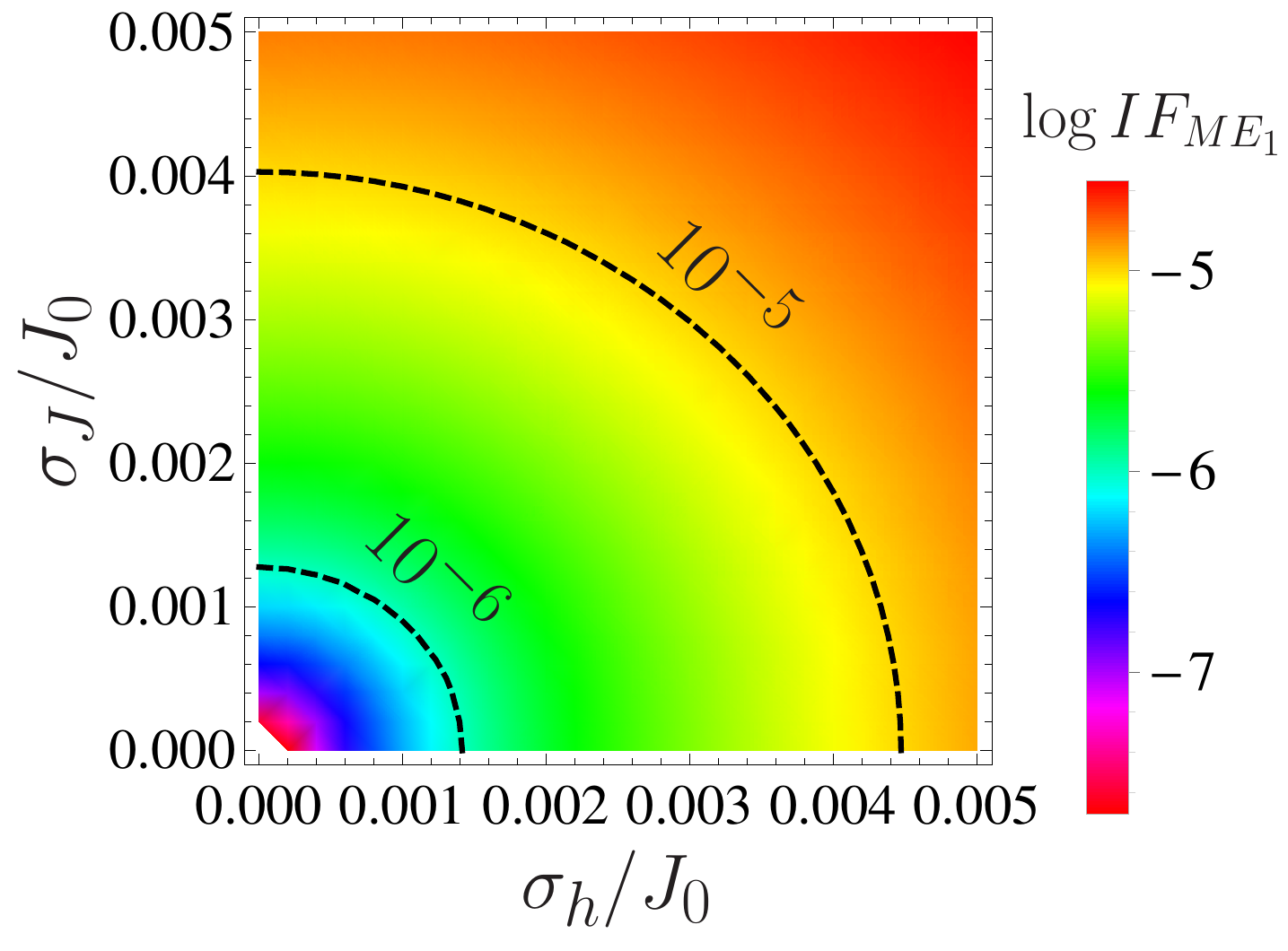}
\includegraphics[width=0.49\columnwidth]{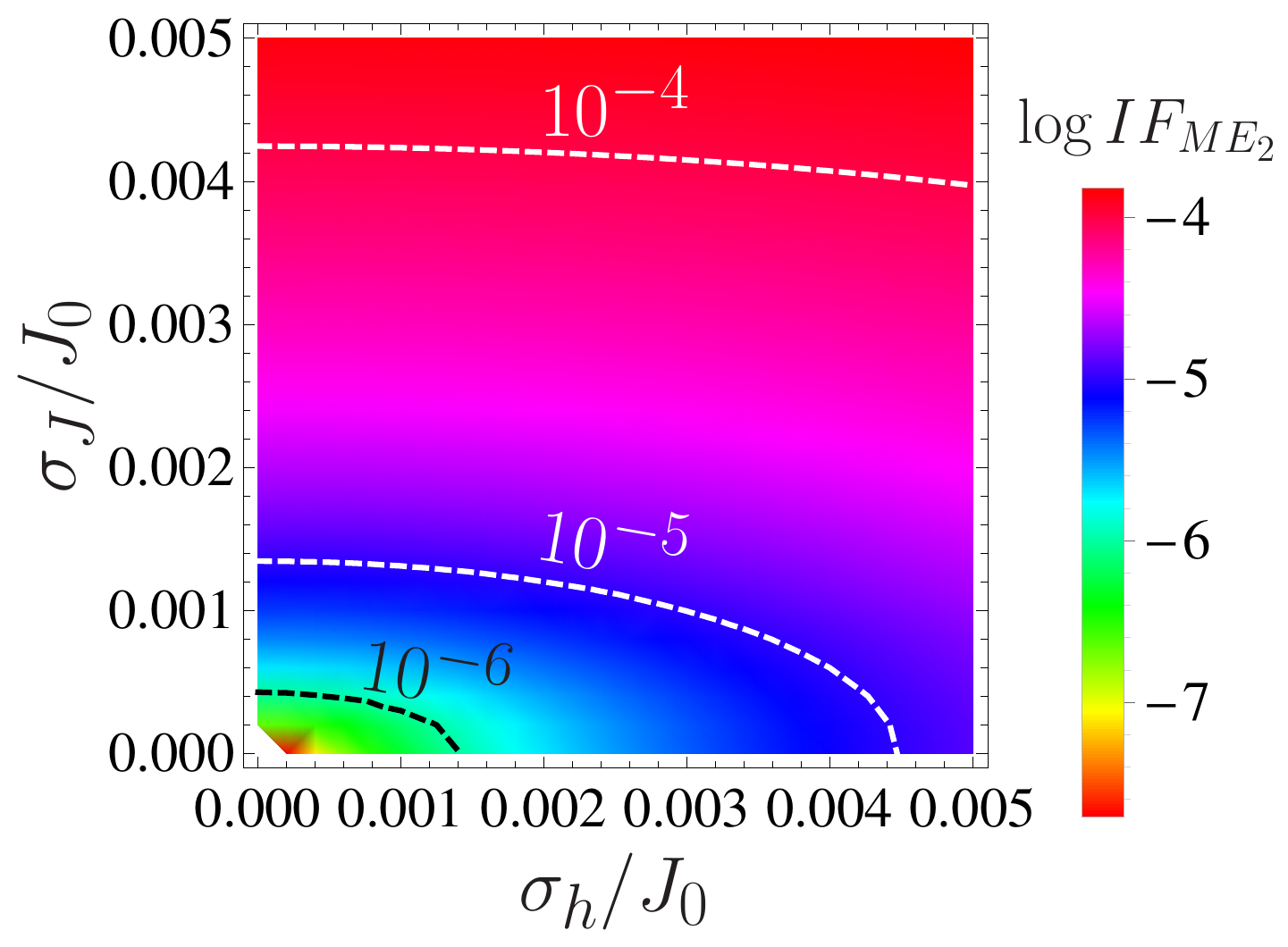}
\caption{Plots of the (common) logarithms of the infidelities, $IF_S=1-F_S$, over the region, $0<\sigma_h/J_0\leq 0.005$ and
$0<\sigma_J/J_0\leq 0.005$.  The labels on the contours are the values of the infidelities along said contours.}\label{Fig:Fidelity_Zoomed_More}
\end{figure}

Finally, we present our results for arbitrary $\sigma_h$ and $\sigma_J$.  We first present a plot of the fidelities
$F_{ME_1}$ and $F_{ME_2}$ over the region, $0<\sigma_h/J_0\leq 2.5$ and $0<\sigma_J/J_0\leq 0.15$ in Fig.\ \ref{Fig:Fidelity_Gen},
as well as ``slices'' of these plots for constant $\sigma_h/J_0$ and constant $\sigma_J/J_0$ in Figs.\ \ref{Fig:Fidelity_ConstSigh}
and \ref{Fig:Fidelity_ConstSigJ}, respectively.  We also indicate experimental values of the disorder, extracted from the
data of Ref.\ \onlinecite{MartinsPRL2016}.  We note that fidelities for realizing the state, $\ket{ME_1}$, in excess of $90\%$
have been achieved, but that those for realizing $\ket{ME_2}$ fall short of this value.  We also see that these values fall short
of the threshold required in order to implement error correction codes; the surface
code error correction schemes with the lowest thresholds require a fidelity of roughly $99\%$ in all operations.  Therefore,
we are interested in finding the regions within which we achieve such high fidelities.  We therefore present plots of both
fidelities over the region, $0<\sigma_h/J_0\leq 0.02$ and $0<\sigma_J/J_0\leq 0.02$, in Fig.\ \ref{Fig:Fidelity_Zoomed},
and over the region, $0<\sigma_h/J_0\leq 0.005$ and $0<\sigma_J/J_0\leq 0.005$, in Fig.\ \ref{Fig:Fidelity_Zoomed_More}.
Because of how close to $1$ the fidelities are, we instead plot the common logarithm of the infidelities $IF_S=1-F_S$ for
visual clarity.  We see that, in both regions, one can already achieve a fidelity within the error correction threshold for values of 
$\sigma_h$ and $\sigma_J$ around $0.02J_0$.  We can
also very clearly see that both types of noise have roughly the same effect on $F_{ME_1}$, but that charge noise has more
of an effect on $F_{ME_2}$ than field noise does.

Overall, our results show that a given magnitude of field noise, as measured by $\sigma_h/J_0$, has less of an effect on
the fidelities for realizing maximally entangled states than an identical amount of charge noise, as measured by $\sigma_J/J_0$.
While they have identical effects on $F_{ME_1}$, we see that the amount of field noise needed to obtain a given effect on
$F_{ME_2}$ is about a factor of $3$ larger than the amount of charge noise needed to see the same effect.
In this sense, we may claim that charge noise is more detrimental to our ability to realize a maximally entangled
state than field noise.  We may break down the effects caused directly by these two types of noise contributing to the
decrease in fidelity as follows.  Charge noise has a single effect, namely, to cause a decay of the amplitude of the oscillations
in the return probability; it has no effect on the steady-state value of the return probability.  Field noise also damps
out the oscillations in the return probability, but a given amount of field noise will have less of an effect on it than
the same amount of charge noise would.  However, field noise has a second effect---it also increases the steady-state
return probability.  This, too, is harmful to the creation of a maximally-entangled state.  Overall, we see that these
effects result in an $F_{ME_1}$ that is just as adversely affected by a given amount of field noise as by the same amount of
charge noise, but $F_{ME_2}$ is affected more strongly by charge noise than by field noise.  However, we see that field
noise affects $F_{ME_2}$ even more strongly than it does the ``quality factor'' $Q$ described earlier.  This shows that
the damping of the oscillations of the return probability is not the only phenomenon that affects the entanglement fidelity---the
fact that the return probability, and thus the initial amplitude of the return probability oscillations, differ from $\tfrac{1}{2}$
also has an effect.

We should note, however, in light of the above discussion that, in actual GaAs experimental setups, the field noise is much
larger in magnitude than charge noise, as can be seen from Fig.\ \ref{Fig:Fidelity_Gen}.  This indicates that, at least in practice,
field noise is a bigger problem for entanglement in GaAs simply because it is the dominant source of noise in the
system.  Even in this case, we may still reduce the effect of field noise by increasing $J_0$---note that the fidelity, like
the return probability and $T_2^{\ast}$, depends on the strength of the two types of noise only in the combinations, $\sigma_h/J_0$
and $\sigma_J/J_0$.  It turns out that, as we tune $J_0$, $\sigma_h$ remains constant, while $\sigma_J$ increases roughly
linearly\cite{BarnesPRB2016}.  This means that, while we cannot reduce the effect of charge noise by simply increasing the
exchange coupling, we can reduce the effect of field noise.  This is especially helpful for GaAs systems since one cannot eliminate
field noise through isotopic purification schemes as one can in Si.  Other things being equal, however, it is obvious that
Si has a great advantage over GaAs simply by virtue of its much lower field noise since Overhauser nuclear fluctuations can
be almost entirely eliminated in Si, but not in GaAs.  On the other hand, the advantage of having much weaker field noise in
Si would be seriously compromised if the charge noise in Si turns out to be stronger than in GaAs.

\section{Discussion}
We have determined the intrinsic coherence time, $T_2^{\ast}$, the steady-state return probability, $P_S$, and the fidelities
$F_{ME_1}$ and $F_{ME_2}$ for producing the maximally entangled states $\ket{ME_1}$ and $\ket{ME_2}$, respectively, starting
from the unentangled state, $\ket{\uparrow\downarrow}$, for a system of two electron spins coupled via the Heisenberg exchange
coupling with both magnetic field and charge noise as a function of the strength of both types of noise.  We employed
the quasistatic bath approximation, taking the distributions for both types of noise to be Gaussian; in the case of charge noise,
we truncated the distribution so that the exchange coupling is always positive.  These results are of direct and immediate relevance
to current quantum computing experiments on coupled electron spins in quantum dots since we have focused on standard experimental observables (return
probability) and used realistic noise models.  In fact, we indicate the values of the disorder that have recently been estimated experimentally
in our presented results\cite{MartinsPRL2016}.  The main goal of our work is to calculate the fidelity of operations that maximally
entangle two electron spins, starting from an unentangled state.  This is an important problem, as universal quantum computation
requires the ability to perform at least one such two-qubit operation, along with arbitrary single-qubit rotations.

We find that, while both types of noise suppress $T_2^{\ast}$, electronic noise has a far greater effect, indicating that
charge noise is much more effective at causing decoherence in the system than field noise.  We also calculate the
steady-state return probability and show that only field noise has any significant effect on it.  While the fact that charge
noise lowers $T_2^{\ast}$ more than field noise would at first seem to indicate that it is much more harmful to
our ability to entangle the electronic spins, the situation is more subtle in the presence of both types of noise.  Because
charge noise does not change the steady-state return probability, our ``na\"ive'' entangling operation still closely approximates
the ideal $\sqrt{\hbox{SWAP}}$ entangling operation as long as it is performed over a time scale much smaller than $T_2^{\ast}$.
This is not the case with field noise, however: any amount of field noise will change the steady-state return probability,
and thus our ``na\"ive'' operation will deviate from the ideal $\sqrt{\hbox{SWAP}}$ entangling operation regardless of how quickly
it is performed.  We thus expect that field noise is more harmful to performing such an operation than our
results for $T_2^{\ast}$ would suggest, and this is in fact borne out in our results for the entanglement fidelities.  We
find that $F_{ME_1}$ is just as greatly reduced by a given magnitude of field noise as by an equal amount of charge noise.
On the other hand, $F_{ME_2}$ is not affected as much by field noise as it is by charge noise, but it is still reduced more
than the ``quality factor'' $Q=e^{-1/J_0 T_2^{\ast}}$ is.  This fact helps to illustrate the effect of the steady-state
return probability shift on the fidelity.  One must, however, take into account the fact that field and charge noise are
typically not of comparable magnitudes in semiconductors, and thus, their actual harmful effects on two-qubit operations
would depend strongly on their actual magnitudes.  This is why we have presented results in this work covering wide parameter
regimes for both field and charge noise strengths.

Our results imply that a given amount of charge noise overall has a more detrimental effect on entanglement fidelity than
an equal amount of field noise; while both types of noise affect the fidelity for producing $\ket{ME_1}$ equally, charge
noise has more of an effect on that for producing $\ket{ME_2}$ than field noise does.  We see, however, that, at least in
GaAs, field noise is much stronger than charge noise, and thus is a bigger problem in this material.  Such noise, however,
can be made much smaller in Si than in GaAs, since isotopic purification can greatly reduce the presence of magnetic isotopes
of Si (${}^{29}$Si, to be exact), while no such reduction is possible in GaAs because the only stable isotopes of Ga and As
have non-zero spin. The same concerns about unavoidable field noise in GaAs arise in P-doped Si as well owing to the fact
that the only stable isotope of P, ${}^{31}$P, has a non-zero spin; however, the field noise will not be as great as in GaAs
since not all of the nuclei present in the sample are magnetic. There is, however, another way to reduce the effects of field
noise that works for both materials.  We can take advantage of the fact that $\sigma_h$ essentially remains constant as one
changes the average exchange coupling $J_0$, while $\sigma_J$ is roughly linear\cite{BarnesPRB2016} in $J_0$.  This means
that, since the fidelities and the return probability depend only on $\sigma_h/J_0$ and $\sigma_J/J_0$, one can reduce the
effects of field noise simply by increasing $J_0$.  In fact, as can be seen from the experimental points indicated in Fig.\
\ref{Fig:Fidelity_Gen}, it is possible to achieve fidelities in excess of $90\%$ for producing the state $\ket{ME_1}$
or close to $90\%$ for producing $\ket{ME_2}$ in current GaAs-based experimental setups by doing this.  Recent experimental
work\cite{NicholArXiv} suggests that a similar approach can be used to suppress charge noise by creating a large magnetic field
gradient across the double quantum dot, provided Bayesian estimation or dynamical decoupling are used simultaneously to mitigate
field noise. Our findings indicate that if both $\sigma_h/J_0$ and $\sigma_J/J_0$ are reduced down to the $1\%$ level, then entanglement
fidelities at or beyond surface code thresholds of $99\%$ can be achieved.

Our work also implies that Si systems in the end are far superior to GaAs in terms of achieving ideal two-qubit gate operations by
virtue of the fact that in Si the field noise can, in principle, be reduced to arbitrarily small values by eliminating background
nuclear spin fluctuations.  Charge noise, on the other hand, is likely to be similar in both systems.  Therefore, Si obviously has
a great advantage over GaAs in terms of noise.  This is not to say that Si has a clear advantage over GaAs; there are challenges
associated with Si as a material platform as well.  For one, the charge carriers in Si have a much higher effective mass than in GaAs.
This means that one is forced to fabricate much smaller dots in order to confine individual electrons.  Another issue is valley
degeneracy of energy levels, which complicates the task of isolating two nondegenerate levels to use as qubit states.  Valley effects
relevant to Si quantum dots have been studied to some extent in the literature\cite{HaoNatCommun2014,SongArXiv2016}.  Our detailed
numerical results presented in this paper provide a quantitative guide on how much both types of noise must be suppressed in experimental
systems for achieving the fidelities above $99\%$ that are necessary for further progress in the field.

\acknowledgements
This work is supported by LPS-MPO-CMTC.


\begin{thebibliography}{99}
\bibitem{ZwanenburgRMP2013}F.\ Zwanenburg, A.\ S.\ Dzurak, A.\ Morello, M.\ Y.\ Simmons, L.\ C.\ L.\ Hollenberg, G.\ Klimeck, S.\ Rogge, S.\ N.\ Coppersmith, and M.\ A.\ Eriksson, \rmp {\bf 85}, 961 (2013).
\bibitem{LossPRA1998}D.\ Loss and D.\ P.\ DiVincenzo, \pra {\bf 57}, 120 (1998).
\bibitem{PlaNature2012}J.\ Pla, K.\ Y.\ Tan, J.\ P.\ Dehollain, W.\ H.\ Lim, J.\ J.\ L.\ Morton, D.\ N.\ Jamieson, A.\ S.\ Dzurak, and A.\ Morello, Nature (London) {\bf 489}, 541 (2012).
\bibitem{PlaNature2013}J.\ Pla, K.\ Y.\ Tan, J.\ P.\ Dehollain, W.\ H.\ Lim, J.\ J.\ L.\ Morton, F.\ A.\ Zwanenburg, D.\ N.\ Jamieson, A.\ S.\ Dzurak and A.\ Morello, Nature (London) {\bf 496}, 334 (2013).
\bibitem{VeldhorstNatNano2014}M.\ Veldhorst, J.\ C.\ C.\ Hwang, C.\ H.\ Yang, A.\ W.\ Leenstra, B.\ de Ronde, J.\ P.\ Dehollain, J.\ T.\ Muhonen, F.\ E.\ Hudson, K.\ M.\ Itoh, A.\ Morello, and A.\ S.\ Dzurak, Nature Nano. {\bf 9}, 981 (2014).
\bibitem{BraakmanNatNano2013}F.\ R.\ Braakman, P.\ Barthelemy, C.\ Reichl, W.\ Wegscheider, and L.\ M.\ K.\ Vandersypen, Nature Nano. {\bf 8}, 432-437 (2013).
\bibitem{OtsukaSciRep2016}T.\ Otsuka, T.\ Nakajima, M.\ R.\ Delbecq, S.\ Amaha, J.\ Yoneda, K.\ Takeda, G.\ Allison, T.\ Ito, R.\ Sugawara, A.\ Noiri, A.\ Ludwig, A.\ D.\ Wieck, and S.\ Tarucha, Sci. Rep. {\bf 6}, 31820 (2016).
\bibitem{ItoArxiv2016}T.\ Ito, T.\ Otsuka, S.\ Amaha, M.\ R.\ Delbecq, T.\ Nakajima, J.\ Yoneda, K.\ Takeda, G.\ Allison, A.\ Noiri, K.\ Kawasaki, and S.\ Tarucha, arXiv:1604.04426.
\bibitem{LevyPRL2002}J.\ Levy, \prl {\bf 89}, 147902 (2002).
\bibitem{PettaScience2005}J.\ R.\ Petta, A.\ C.\ Johnson, J.\ M.\ Taylor, E.\ A.\ Laird, A.\ Yacoby, M.\ D.\ Lukin, C.\ M.\ Marcus, M.\ P.\ Hanson, A.\ C.\ Gossard, Science {\bf 309}, 2180 (2005).
\bibitem{FolettiNatPhys2009}S.\ Foletti, H.\ Bluhm, D.\ Mahal, V.\ Umansky, and A.\ Yacoby, Nature Phys. {\bf 5}, 903 (2009).
\bibitem{MauneNature2012}B.\ Maune, M.\ G.\ Borselli, B.\ Huang, T.\ D.\ Ladd, P.\ W.\ Deelman, K.\ S.\ Holabird, A.\ A.\ Kiselev, I.\ Alvarado-Rodriguez, R.\ S.\ Ross, A.\ E.\ Schmitz, M.\ Sokolich, C.\ A.\ Watson, M.\ F.\ Gyure, and A.\ T.\ Hunter, Nature (London) {\bf 481}, 344 (2012).
\bibitem{ShulmanNatCommun2014}M.\ Shulman, S.\ P.\ Harvey, J.\ M.\ Nichol, S.\ D.\ Bartlett, A.\ C.\ Doherty, V.\ Umansky, and A.\ Yacoby, Nature Commun. {\bf 5}, 5156 (2014).
\bibitem{DialPRL2013}O.\ E.\ Dial, M.\ D.\ Shulman, S.\ P.\ Harvey, H.\ Bluhm, V.\ Umansky, and A.\ Yacoby, \prl {\bf 110}, 146804 (2013).
\bibitem{MartinsPRL2016}F.\ Martins, F.\ K.\ Malinowski, P.\ D.\ Nissen, E.\ Barnes, S.\ Fallahi, G.\ C.\ Gardner, M.\ J.\ Manfra, C.\ M.\ Marcus, and F.\ Kuemmeth, \prl {\bf 116}, 116801 (2016).
\bibitem{DiVincenzoNature2000}D.\ P.\ DiVincenzo, D.\ Bacon, J.\ Kempe, G.\ Burkard, and K.\ B.\ Whaley, Nature (London) {\bf 408}, 339 (2000).
\bibitem{MedfordNatNano2013}J.\ Medford, J.\ Beil, J.\ M.\ Taylor, S.\ D.\ Bartlett, A.\ C.\ Doherty, E.\ I.\ Rashba, D.\ P.\ DiVincenzo, H.\ Lu, A.\ C.\ Gossard, and C.\ M.\ Marcus, Nature Nano. {\bf 8}, 654 (2013).
\bibitem{MedfordPRL2013}J.\ Medford, J.\ Beil, J.\ M.\ Taylor, E.\ I.\ Rashba, H.\ Lu, A.\ C.\ Gossard, and C.\ M.\ Marcus, \prl {\bf 111}, 050501 (2013).
\bibitem{EngSciAdv2015}K.\ Eng, T.\ D.\ Ladd, A.\ Smith, M.\ G.\ Borselli, A.\ A.\ Kiselev, B.\ H.\ Fong, K.\ S.\ Holabird, T.\ M.\ Hazard, B.\ Huang, P.\ W.\ Deelman, I.\ Milosavljevic, A.\ E.\ Schmitz, R.\ S.\ Ross, M.\ F.\ Gyure, and A.\ T.\ Hunter, Sci. Adv. {\bf 1}, e150021 (2015).
\bibitem{ShimPRB2016}Y.-P.\ Shim and C.\ Tahan, \prb {\bf 93}, 121410 (2016).
\bibitem{ShiPRL2012}Z.\ Shi, C.\ B.\ Simmons, J.\ R.\ Prance, J.\ K.\ Gamble, T.\ S.\ Koh, Y.\-P.\ Shim, X.\ Hu, D.\ E.\ Savage, M.\ G.\ Lagally, M.\ A.\ Eriksson, M.\ Friesen, and S.\ N.\ Coppersmith, Phys. Rev. Lett. {\bf 108}, 140503 (2012).
\bibitem{KimNature2014}D.\ Kim, Z.\ Shi, C.\ B.\ Simmons, D.\ R.\ Ward, J.\ R.\ Prance, T.\ S.\ Koh, J.\ K.\ Gamble, D.\ E.\ Savage, M.\ G.\ Lagally, M.\ Friesen, S.\ N.\ Coppersmith, and M.\ A.\ Eriksson, Nature (London) {\bf 511}, 70 (2014).
\bibitem{KimnpjQI2015}D.\ Kim, D.\ R.\ Ward, C.\ B.\ Simmons, D.\ E.\ Savage, M.\ G.\ Lagally, M.\ Friesen, S.\ N.\ Coppersmith, and M.\ A.\ Eriksson, npj Quant. Inf. {\bf 1}, 15004 (2015).
\bibitem{vanWeperenPRL2011}I.\ van Weperen, B.\ D.\ Armstrong, E.\ A.\ Laird, J.\ Medford, C.\ M.\ Marcus, M.\ P.\ Hanson, and A.\ C.\ Gossard, \prl {\bf 107}, 030506 (2011).
\bibitem{VeldhorstNature2016}M.\ Veldhorst, C.\ H.\ Yang, J.\ C.\ C.\ Hwang, W.\ Huang, J.\ P.\ Dehollain, J.\ T.\ Muhonen, S.\ Simmons, A.\ Laucht, F.\ E.\ Hudson, K.\ M.\ Itoh, A.\ Morello, and A.\ S.\ Dzurak, Nature {\bf 526}, 410 (2016).
\bibitem{NowackScience2011}K.\ Nowack, M.\ Shafiei, M.\ Laforest, G.\ E.\ D.\ K.\ Prawiroatmodjo, L.\ R.\ Schreiber, C.\ Reichl, W.\ Wegscheider, and L.\ M.\ K.\ Vandersypen, Science {\bf 333}, 1269 (2011).
\bibitem{ShulmanScience2012}M.\ Shulman, O.\ E.\ Dial, S.\ P.\ Harvey, H.\ Bluhm, V.\ Umansky, and A.\ Yacoby, Science {\bf 336}, 202 (2012).
\bibitem{NicholArXiv}J.\ M.\ Nichol, L.\ A.\ Orona, S.\ P.\ Harvey, S.\ Fallahi, G.\ C.\ Gardner, M.\ J.\ Manfra, and A.\ Yacoby, arXiv:1608.04258.
\bibitem{FowlerPRA2012}A.\ G.\ Fowler, M.\ Mariantoni, J.\ M.\ Martinis, and A.\ N.\ Cleland, \pra {\bf 86}, 032324 (2012).
\bibitem{DeSousaPRB2003}R.\ de Sousa and S.\ Das Sarma, \prb {\bf 67}, 033301 (2003).
\bibitem{HuPRL2005}X.\ Hu and S.\ Das Sarma, \prl {\bf 96}, 100501 (2006).
\bibitem{ViolaPRA1998}L.\ Viola and S.\ Lloyd, Phys. Rev. A {\bf 58}, 2733 (1998).
\bibitem{CywinskiPRB2008}\L.\ Cywinski, R.\ M.\ Lutchyn, C.\ P.\ Nave, and S.\ Das Sarma, Phys. Rev. B {\bf 77}, 174509 (2008).
\bibitem{BluhmPRL2010}H.\ Bluhm, S.\ Foletti, D.\ Mahalu, V.\ Umansky, and A.\ Yacoby, \prl {\bf 105}, 216803 (2010).
\bibitem{MalinkowskiArXiv}F.\ K.\ Malinowski, F.\ Martins, P.\ D.\ Nissen, E.\ Barnes, \L.\ Cywi\'nski, M.\ S.\ Rudner, S.\ Fallahi, G.\ C.\ Gardner, M.\ J.\ Manfra, C.\ M.\ Marcus, F.\ Kuemmeth, Nat. Nanotechnol. {\bf 12}, 16 (2017).
\bibitem{SergeevichPRA2011}A.\ Sergeevich, A.\ Chandran, J.\ Combes, S.\ D.\ Bartlett, and H.\ M.\ Wiseman, \pra {\bf 84} 052315 (2011).
\bibitem{WitzelPRL2010}W.\ M.\ Witzel, M.\ S.\ Carroll, A.\ Morello, \L.\ Cywi\'nski, and S. Das Sarma, \prl {\bf 105}, 187602 (2010).
\bibitem{WangNatCommun2012}X.\ Wang, L.\ S.\ Bishop, J.\ P.\ Kestner, E.\ Barnes, K.\ Sun, and S.\ Das Sarma, Nature Commun. {\bf 3}, 997 (2012).
\bibitem{KestnerPRL2013}J.\ P.\ Kestner, X.\ Wang, L.\ S.\ Bishop, E.\ Barnes, and S.\ Das Sarma, \prl {\bf 110}, 140502 (2013).
\bibitem{WangPRA2014}X.\ Wang, L.\ S.\ Bishop, E.\ Barnes, J.\ P.\ Kestner, and S.\ Das Sarma, \pra {\bf 89}, 022310 (2014).
\bibitem{KhodjastehPRA2012}K.\ Khodjasteh, H.\ Bluhm, and L.\ Viola, \pra {\bf 86}, 042329 (2012).
\bibitem{BarnesSciRep2015}E.\ Barnes, X.\ Wang, and S.\ Das Sarma, Sci. Rep. {\bf 5}, 12685 (2015).
\bibitem{UhrigPRL2007}G.\ S.\ Uhrig, \prl {\bf 98}, 100504 (2007). 
\bibitem{WitzelPRL2007}W.\ M.\ Witzel and S.\ Das Sarma, \prl {\bf 98}, 077601 (2007).
\bibitem{LeePRL2008}B.\ Lee, W.\ M.\ Witzel, and S.\ Das Sarma, \prl {\bf 100}, 160505 (2008).
\bibitem{YaoPRL2007}W.\ Yao, R.\-B.\ Liu, and L.\ J.\ Sham, \prl {\bf 98}, 077602 (2007).
\bibitem{YuPRB2003}T.\ Yu and J.\ H.\ Eberly, \prb {\bf 68}, 165322 (2003).
\bibitem{AnnPRB2007}K.\ Ann and G.\ Jaeger, \prb {\bf 75} 115307 (2007).
\bibitem{DePRB2011}A.\ De, A.\ Lang, D.\ Zhou, and R.\ Joynt, \pra {\bf 83}, 042331 (2011).
\bibitem{BragarPRB2015}I.\ Bragar, and \L.\ Cywi\'nski, \prb {\bf 91}, 155310 (2015).
\bibitem{SzankowskiQIP2015}P.\ Sza\'nkowski, M.\ Trippenbach, \L.\ Cywi\'nski, and Y.\ B.\ Band, Quantum Info. Process. {\bf 14}, 3367 (2015).
\bibitem{CoishPRB2005}W.\ A.\ Coish and D.\ Loss, \prb {\bf 72}, 125337 (2005).
\bibitem{KlauserPRB2006}D.\ Klauser, W.\ A.\ Coish, and D.\ Loss, \prb {\bf 73}, 205302 (2006).
\bibitem{BarnesPRB2016}E.\ Barnes, M.\ S.\ Rudner, F.\ Martins, F.\ K.\ Malinowski, C.\ M.\ Marcus, and F.\ Kuemmeth, \prb {\bf 93}, 121407R (2016).
\bibitem{DasSarmaPRB2016}S.\ Das Sarma, R.\ E.\ Throckmorton, and Y.-L.\ Wu, \prb {\bf 94}, 045435 (2016).
\bibitem{CywinskiAPP2011}\L.\ Cywi\'nski, Acta Phys.\ Pol.\ {\bf 119}, 576 (2011).
\bibitem{NederPRB2011}I.\ Neder, M.\ S.\ Rudner, H.\ Bluhm, S.\ Foletti, B.\ I.\ Halperin, and A.\ Yacoby, \prb {\bf 84}, 035441 (2011).
\bibitem{MedfordPRL2012}J.\ Medford, \L.\ Cywi\'nski, C.\ Barthel, C.\ M.\ Marcus, M.\ P.\ Hanson, and A.\ C.\ Gossard, \prl {\bf 108}, 086802 (2012).
\bibitem{NielsenBook}M.\ A.\ Nielsen and I.\ L.\ Chuang, {\it Quantum Computation and Quantum Information} (Cambridge University Press, Cambridge, 2000).
\bibitem{MagesanPRA2012}E.\ Magesan, J.\ M.\ Gambetta, and J.\ Emerson, \pra {\bf 85}, 042311 (2012).
\bibitem{LaddPRB2012}T.\ D.\ Ladd, \prb {\bf 86}, 125408 (2012)
\bibitem{HaoNatCommun2014}X.\ Hao, R.\ Ruskov, M.\ Xiao, C.\ Tahan, and H.-W. Jiang, Nature Commun. {\bf 5}, 3860 (2014).
\bibitem{SongArXiv2016}Y.\ Song and S.\ Das Sarma, Appl.\ Phys.\ Lett.\ {\bf 109}, 253113 (2016).
\end{thebibliography}
\end{document}